\documentclass[showpacs,amsmath,amssymb, nobibnotes, aps, prl,showkeys, floatfix]{revtex4}
\usepackage{graphicx}
\usepackage{dcolumn}
\usepackage{bm}
\usepackage{docs}
\usepackage{amssymb}
\usepackage{amsmath}
\usepackage{epsfig}
\expandafter\ifx\csname package@font\endcsname\relax\else
 \expandafter\expandafter
 \expandafter\usepackage
 \expandafter\expandafter
 \expandafter{\csname package@font\endcsname}%
\fi

\newcommand{\ltsima} {$\; \buildrel < \over \sim \;$}
\newcommand{\gtsima} {$\; \buildrel > \over \sim \;$}
\newcommand{\lta} {\lower.5ex\hbox{\ltsima}}
\newcommand{\gta} {\lower.5ex\hbox{\gtsima}}

\newcommand{\lsim}{\raisebox{-.4ex}{$\stackrel{<}{\scriptstyle \sim}$}}
\newcommand{\gsim}{\raisebox{-.4ex}{$\stackrel{>}{\scriptstyle \sim}$}}

\newcommand{\RNum}[1]{\uppercase\expandafter{\romannumeral #1\relax}}

\begin{document}

\title[Newtonian analogues description of naked singularities]{
Newtonian analogue of static general relativistic spacetimes: An extension to naked singularities}  

\author {Shubhrangshu Ghosh$^{1}$, Tamal Sarkar$^{1,2}$, Arunava Bhadra$^{1}$ } 

\affiliation{ $^{1}$ High Energy $\&$ Cosmic Ray Research Center, University of North Bengal, Post N.B.U, Siliguri 734013, India. }
\affiliation{ $^{2}$ University Science Instrumentation Center, University of North Bengal, Post N.B.U, Siliguri 734013, India. }

\begin{abstract}

We formulate a generic Newtonian like analogous potential for static spherically symmetric general relativistic (GR) spacetime, and subsequently derived proper Newtonian like analogous potential corresponding to Janis-Newman-Winicour (JNW) and Reissner-Nordstr\"{o}m (RN) spacetimes, both exhibiting naked singularities. The derived potentials found to reproduce the entire GR features including the orbital dynamics of the test particle motion and the orbital trajectories, with precise accuracy. The nature of the particle orbital dynamics including their trajectory profiles in JNW and RN geometries show altogether different behavior with distinctive traits as compared to the nature of particle dynamics in Schwarzschild geometry. Exploiting the Newtonian like analogous potentials, we found that the radiative efficiency of a geometrically thin and optically thick Keplerian accretion disk around naked singularities corresponding to both JNW and RN geometries, in general, is always higher than that for Schwarzschild geometry. The derived potentials would thus be useful to study astrophysical processes, especially to investigate more complex accretion phenomena in AGNs or in XRBs in the presence of naked singularities and thereby exploring any noticeable differences in their observational features from those in the presence of BHs to ascertain outstanding debatable issues relating to gravity - whether the end state of gravitational collapse in our physical Universe renders black hole (BH) or naked singularity.

\end{abstract}

\maketitle

\section{Introduction}

An inevitable feature of exact solutions in general relativity under different general physical conditions is the occurrence of singularities. Whether such singularities will always be covered (by the event horizon) to distant observers or not is an interesting unresolved question. The {\it cosmic censorship} conjecture [1] prevents the development of a naked singularity generically in realistic gravitational collapse. However, the question of cosmic censorship is still open due to lack of any rigorous proof of the conjecture. 
 
If naked singularities exist in nature as real astrophysical objects it is worthwhile to explore whether or not the naked singularities give different observational predictions than those due to black holes (BHs) which could be utilized to discriminate these two objects observationally. Gravitational lensing by naked singularities, particularly in strong field regime, is found to have some interesting characteristics difference from those by BHs [2]. The time delay between successive relativistic images in gravitational lensing also exhibit different behavior for naked singularity and BH lens. Even the time delays of relativistic images are found negative for strongly naked singularity solution [3]. The fluxes of escaping particles produced in ultra high energy collision of particles in the vicinity of a naked singularity also bear the characteristics of naked singularity [4]. The properties of stable circular orbits around naked singularity solution are significantly different than those around a Schwarzschild BH with the same mass and thus the accretion disk around the naked singularity could be observationally distinguished from that around a BH [5]. 

It has long been argued innumerable times in the literature about the necessity of using pseudo-Newtonian potentials (PNPs) [6-8] to study complex accretion phenomena around BHs/compact object, ever since the seminal work of Paczy\'nski and Witta [9]. Series of PNPs or Newtonian like analogous potentials of the corresponding relativistic geometries till date exist in the literature [6-20] which, however, are mostly proposed for corresponding BH solutions. In this work we construct PNPs for the naked singularity solutions, corresponding to two `static non vacuum solutions' to Einstein's field equations: the Janis-Newman-Winicour (JNW) metric [21] which is an exact solution of GR field equations in presence of a minimally coupled massless scalar field exhibiting naked singularity and the Reissner-Nordstr\"{o}m (RN) solution, which is the well known unique asymptotically flat solution of the Einstein-Maxwell equations describing a charged nonrotating metric that also exhibits naked singularity for certain choices of solution parameters. We formulate the corresponding PNPs proceeding directly from the conserved Hamiltonian of the system following [20,7,8]. We refer these PNPs as Newtonian like analogous potentials (NAP). 

Note that NAP for the RN geometry could also be obtained from a generic expression given in [22] but the formulation could not be applicable to JNW spacetime. Therefore we first deduce a proper potential analogue in the Newtonian framework corresponding to a most generalized form of static GR spacetime which is therefore applicable to JNW metric. We study the complete orbital dynamics of the test particle motion exploiting NAPs for JNW and RN geometries and compare them with the results for the Schwarzschild geometry as well as among themselves. We also study extensively the general orbital trajectory profiles for both these geometries, in the modified Newtonian analogue. Apart from the use of these potentials to investigate the dynamical nature of accretion flows, the Newtonian analogous potentials could be comprehensively used for generic astrophysical purposes relevant to JNW and RN geometries. In recent times, some properties of the circular orbit dynamics of test particle motion have been investigated in both JNW and RN spacetimes [23,24,25] in GR framework which can be reproduced employing relevant NAPs.

In the next section, we derive a generic Newtonian analogous potential for a general class of static GR spacetime. Subsequently in \S \RNum{3} and \S \RNum{4}, we analyze particle trajectories for the JNW and RN metrics, respectively, in the Newtonian analogous framework, laying emphasis on circular orbital dynamics. In \S \RNum{5}, we investigate the behavior of the trajectory profiles of the test particle corresponding to both JNW and RN geometries comprehensively, with a comparison among themselves and that with the Schwarzschild case. In \S \RNum{6}, we apply the NAPs corresponding to JNW and RN geometries to analyze a simplistic accretion flow system. In \S \RNum{7} we furnish a general discussion on our methods and results. Finally, we culminate in \S \RNum{8} with a summary and conclusion. 

\section{Formulation of a Newtonian analogous potential corresponding to the most general static GR spacetime}

In general relativity, static spacetimes are among the simplest class of Lorentzian manifolds with a non-vanishing timelike irrotational Killing vector field $K^{\alpha}$. As we intend to study both JNW and RN metrics, in standard spherical coordinates system, we choose to write the static GR spacetime represented by the form 
\begin{eqnarray}
ds^2 =-f(r)^\beta \, c^2 \, dt^2 + \frac{1}{f(r)^\beta} \, dr^2 + f(r)^{1-\beta} r^2 d\Omega^2 \, , 
\label{1}
\end{eqnarray}
where f(r) is the generic metric function and $\beta$ is an arbitrary constant. $d\Omega^2 = d\theta^2 +  \sin^2 \theta \, d\phi^2 $. With $\beta =1$, $ds^2$ would reduce to the usual static geometries like Schwarzschild or Schwarzschild de-Sitter or RN with suitable choice of f(r). The Lagrangian density of the particle of mass $m$ in this spacetime is then given by
\begin{eqnarray}
2 {\cal L}=-f(r)^\beta \, c^2 \, \left(\frac{dt}{d\tau} \right)^2 + \frac{1}{f(r)^\beta}\, \left(\frac{dr}{d\tau}\right) ^2 + f(r)^{1-\beta} \, 
r^2 \, \left(\frac{d\Omega}{d\tau}\right) ^2\, .  
\label{2}
\end{eqnarray}
From the symmetries, the two constants of motion corresponding to two ignorable coordinates $t$ and $\Omega$ are given by
\begin{eqnarray}
{\cal P}_t =\frac{\partial \cal L}{\partial \tilde t}= - c^2 \, f(r)^\beta \, \frac{dt}{d\tau}={\rm constant}=-\epsilon 
\label{3}
\end{eqnarray}               
and 
\begin{eqnarray}
 {\cal P}_{\Omega}=\frac{\partial \cal L}{\partial \tilde \Omega} = 
r^2 f(r)^{1-\beta} \, \frac{d\Omega}{d\tau}={\rm constant} = \lambda \, .
\label{4}
\end{eqnarray}
where $\epsilon$ and $\lambda$ are specific energy and specific generalized angular momentum of the orbiting particle respectively. Here $\tilde t = {dt}/d{\tau}$ and $\tilde \Omega ={d\Omega}/d{\tau}$, the derivatives with respect to the proper time $\tau$. Using Eq. (3) we can write  
\begin{eqnarray}
\frac{dt}{d\tau} = \frac{\epsilon}{c^2} \frac{1}{f(r)^\beta} \, .
\label{5}
\end{eqnarray}
We can write ${\cal L}$, given by 
\begin{eqnarray}
2 {\cal L} =g_{\rm {\nu \mu}} \, p^{\nu} \, p^{\mu} =-m^2 c^2 \, ,
\label{6}
\end{eqnarray}
which is itself a constant in the local inertial frame. Using the above 
equations, we obtain 
\begin{eqnarray}
\left (\frac{dr}{d\tau} \right)^{2} = \left(\frac{\epsilon^{2}}{c^{2}}- c^{2} \right) 
-c^{2} \, \left(f(r)^\beta-1 \right) - f(r)^{2 \beta-1} \, \frac{\lambda^2}{r^2}  \, .
\label{7}
\end{eqnarray}
We define $E_{\rm GN} = (\epsilon^2 - c^4)/{2c^2}$ in the local inertial frame of the test particle motion, which is also the conserved Hamiltonian of the system (see [8-9] for a discussion). `$\rm GN$' symbolizes `GR-Newtonian'. In the low energy limit of the test particle motion, i.e., $\epsilon/c^2 \sim 1$, using Eqs. (5) and (6), we obtain ${dr}/{dt}$ as given by  
\begin{eqnarray}
\frac{dr}{dt} = f(r)^\beta \, \sqrt{2 \, E_{\rm GN} 
-c^2 \, \left(f(r)^\beta -1 \right) - \, {\dot \Omega}^2 \frac{r^2}{f(r)^{2 \gamma-1}} }\, ,
\label{8}
\end{eqnarray}
where, $\dot \Omega= {d\Omega}/{dt} = f(r)^{2 \beta -1} \, {\lambda}/{r^2}$ is the derivative with respect to coordinate time `$t$'. In the low energy limit of the test particle motion which is our necessary condition for the potential formulation, we write the generalized conserved Hamiltonian using Eq. (8), given by 
\begin{eqnarray}
E_{\rm GN} = \frac{1}{2}\left(\frac{\dot{r}^{2}}{f^{2\beta}}+\frac{r^{2} \, \dot{\Omega}^{2}}{f^{2\beta-1}}\right)+\frac{c^{2}}{2}\,\left(f^{\beta}-1\right) \, .
\label{9}
\end{eqnarray}
Thus the generalized Hamiltonian $E_{\rm GN}$ in the low energy limit should be equivalent to the Hamiltonian in Newtonian regime. In the spherical polar geometry, the Hamiltonian in the Newtonian regime with a generalized potential will be equivalent to Eq. (9), as given by,  
\begin{eqnarray}
E_{\rm GN} \equiv  \frac{1}{2} \left({\dot r}^2 + r^2 {\dot \Omega}^2 \right) 
+ V_{\rm GN} - {\dot r} \frac{\partial V_{\rm GN}}{\partial {\dot r}} 
- {\dot \Omega} \frac{\partial V_{\rm GN}}{\partial {\dot \Omega}} \, , 
\label{10}
\end{eqnarray}
where $T = 1/2 \, \left({\dot r}^2 + r^2 {\dot \Omega}^2 \right)$ is the non-relativistic specific kinetic energy of the test particle. $V_{\rm GN}$ in Eq. (10) is the Newtonian analogous potential which would then be given by  
\begin{eqnarray}
V_{\rm GN}  =  \frac{c^{2}(f^{\beta}-1)}{2}-\left(\frac{1-f^{2\beta-1}}{2 \, f^{2\beta-1}}\right)\left[\frac{f^{2\beta}-1}{f \, \left(f^{2\beta-1}-1\right)}\dot{\, r}^{2}+r^{2}\,\dot{\Omega}^{2}\right] \, .
\label{11}
\end{eqnarray}
Overdots here always denote the derivative with respect to coordinate time `$t$'. Thus, $V_{\rm GN}$ in  Eq. (11) is  
the most generalized three dimensional potential in spherical geometry in the modified Newtonian analogue corresponding to any generalized static GR metric in the form given by Eq. (1), in the low energy limit of the test particle motion. The first term of the Newtonian analogous potential contains the explicit information of the source along with the extra field coupled with the curvature. Without any contribution from the external coupling terms, this term reproduces the classical Newtonian gravity for a purely spherically symmetric mass distribution of the source. The second term contains the explicit information of the test particle velocity, accountable for the approximate relativistic modification of the classical Newtonian gravity. In the next two sections we will analyze complete orbital dynamics of the test particle motion in the gravitational field of two generalized non vacuum static spacetimes: JNW and RN geometries, in the Newtonian analogous framework.

\section {Orbital dynamics around JNW spacetime}

We first analyze the test particle dynamics around JNW geometry in the modified Newtonian analogue and compare that with the corresponding GR results. JNW metric is the static spherically symmetric solution of the GR field equations in the presence of a minimally coupled scalar field. It exhibits naked singularity having a constant scalar charge $q$. For JNW geometry, the arbitrary constant in Eq. (1) $\beta=\gamma$, and the the metric function of JNW geometry is $f(r)= 1- \frac{2r_s}{\gamma r}$, where $r_s = {GM}/{c^2}$ and $\gamma$ is a constant parameter. The scalar field is  
$\varphi = \sqrt{\frac{2(1-\gamma^{2})}{16\pi}} {\rm ln} \left(1-\frac{2GM}{\gamma c^2 r}\right) $. The real scalar field demands $0 < \gamma \leq 1$ 
where $\gamma = \frac{1}{\sqrt{1+q^2/r^2_s}}$. The metric function corresponding to JNW geometry diverges at $\gamma r - 2r_s = 0$ exhibiting naked singularity when $\gamma \ne 1$. Using the relation of $f(r)$, and $\gamma$, the three dimensional generalized potential in spherical geometry in Newtonian analogue corresponding to JNW geometry in the low energy limit is obtained using Eq. (11), given by
\begin{eqnarray}
V_{\rm JNW} = \frac{c^2}{2} \left[\left(1-\frac{2r_s}{\gamma r}\right)^\gamma - 1\right]
- \left[\frac{\left(\gamma r \right)^{2\gamma-1} - \left(\gamma r - 2r_s \right)^{2\gamma-1} }{2\left(\gamma r - 2r_s \right)^{2\gamma-1} } \right] 
\left[\frac{\left(\gamma r - 2r_s \right)^{2\gamma} - (\gamma r)^{2\gamma} }
{ (\gamma r - 2 r_s) \left[\left(\gamma r - 2r_s \right)^{2\gamma-1} 
- (\gamma r)^{2\gamma-1} \right] } \, {\dot r}^2 + r^2 \dot \Omega^2 \right] \, ,
\label{12}
\end{eqnarray}
The Newtonian analogous potential in Eq. (12) would be referred to as JNW analogous potential. The timelike circular geodesics which we would be interested in are possible only for $r > 2{r_s}/\gamma$. With $\gamma=1$, we recover the usual Schwarzschild BH solution and the corresponding Newtonian analogous potential. 
The Lagrangian per unit mass corresponding to this potential is given by, 
\begin{eqnarray}
{\cal L}_{\rm JNW} = \frac{(\gamma r)^{2\gamma -1}}{2} \left[\frac{\gamma r \, \dot r ^{2}}{(\gamma r - 2r_s)^{2\gamma}} + \frac{r^2 \dot \Omega^{2}}{(\gamma r - 2r_s)^{2\gamma-1} } \right] - \frac{c^2}{2} \left[\left(1-\frac{2r_s}{\gamma  r}\right)^\gamma - 1\right]  \, ,
\label{13}
\end{eqnarray}
where ${\dot \Omega}^2= {\dot \theta}^2 + \sin^2 \theta \, {\dot \phi}^2$. We next compute the conserved specific angular momentum and specific Hamiltonian using $V_{\rm JNW}$,  which are given by
\begin{eqnarray}
 \lambda_{\rm JNW} = \frac{(\gamma r)^{2\gamma -1} r^2 \, \dot \Omega}{(\gamma r - 2r_s)^{2\gamma -1} }  
\label{14}
\end{eqnarray}
and 
\begin{eqnarray}
 E_{\rm JNW} =\frac{(\gamma r)^{2\gamma -1}}{2} \left[\frac{\gamma r \, \dot r ^{2}}{(\gamma r - 2r_s)^{2\gamma}} + \frac{r^2 \dot \Omega^{2}}{(\gamma r - 2r_s)^{2\gamma-1} } \right] + \frac{c^2}{2} \left[\left(1-\frac{2r_s}{\gamma  r}\right)^\gamma - 1\right]  \, ,
\label{15}
\end{eqnarray}
respectively.  
Using Eqs. (14) and (15), $\dot r$ is given by
\begin{eqnarray}
\frac{dr}{dt} = \left(1-\frac{2r_s}{\gamma r} \right)^\gamma 
\sqrt{2 E_{\rm JNW}  - c^2 \frac{(\gamma r - 2r_s)^{\gamma} - (\gamma r)^{\gamma}}{(\gamma r)^\gamma } - \frac{(\gamma r - 2r_s)^{2\gamma-1}}{(\gamma r)^{2\gamma-1}} \, \frac{\lambda^2_{\rm JNW}}{r^2}  }  \, , 
\label{16}
\end{eqnarray}
which is exactly equivalent to $\dot r$ in general relativity in low energy limit. Next we furnish JNW analogous potential 
in terms of conserved Hamiltonian $E_{\rm JNW}$ and angular momentum $\lambda_{\rm JNW}$, given by   
\begin{eqnarray}
V_{\rm JNW}  = \frac{c^2}{2} \left[\left(1-\frac{2r_s}{\gamma r}\right)^\gamma - 1\right]
- \frac{1}{2} \left[1 - \left(1-\frac{2r_s}{\gamma r}\right)^{2\gamma -1} \right] 
\left[ \frac{\lambda^2_{\rm JNW}}{r^2} 
\frac{(\gamma r - 2r_s)^{2\gamma -1}}{(\gamma r)^{2\gamma -1}} 
\left(1 - \frac{1}{\gamma r} \, \frac{(\gamma r)^{2\gamma}-(\gamma r -2r_s)^{2\gamma}}{(\gamma r )^{2\gamma -1} -  (\gamma r - 2r_s)^{2\gamma -1} } \right)  \right]  \nonumber \\
- \frac{1}{2} \left[1 - \left(1-\frac{2r_s}{\gamma r}\right)^{2\gamma} \right]\, \left[ 2 E_{\rm JNW}  - c^2 \frac{(\gamma r - 2r_s)^{\gamma} - (\gamma r)^{\gamma}}{(\gamma r)^\gamma } \right] \, . \nonumber \\ 
\label{17}
\end{eqnarray}
In Fig. 1 we depict the radial profiles of $V_{\rm JNW}$ for different $\lambda_{\rm JNW}$ corresponding to different $\gamma$ considering low energy limit as well as semi-relativistic energy of the test particle motion. With the increase in $\lambda_{\rm JNW}$ and simultaneously with the decrease in $\gamma$ (i.e. as one departs more from Schwarzschild BH solution), the nature of the profiles of $V_{\rm JNW}$ shows contrasting behavior as compared to the scenario in the Schwarzschild case.


\begin{figure}
\begin{center}
\includegraphics[width = 0.95\textwidth,angle=0]{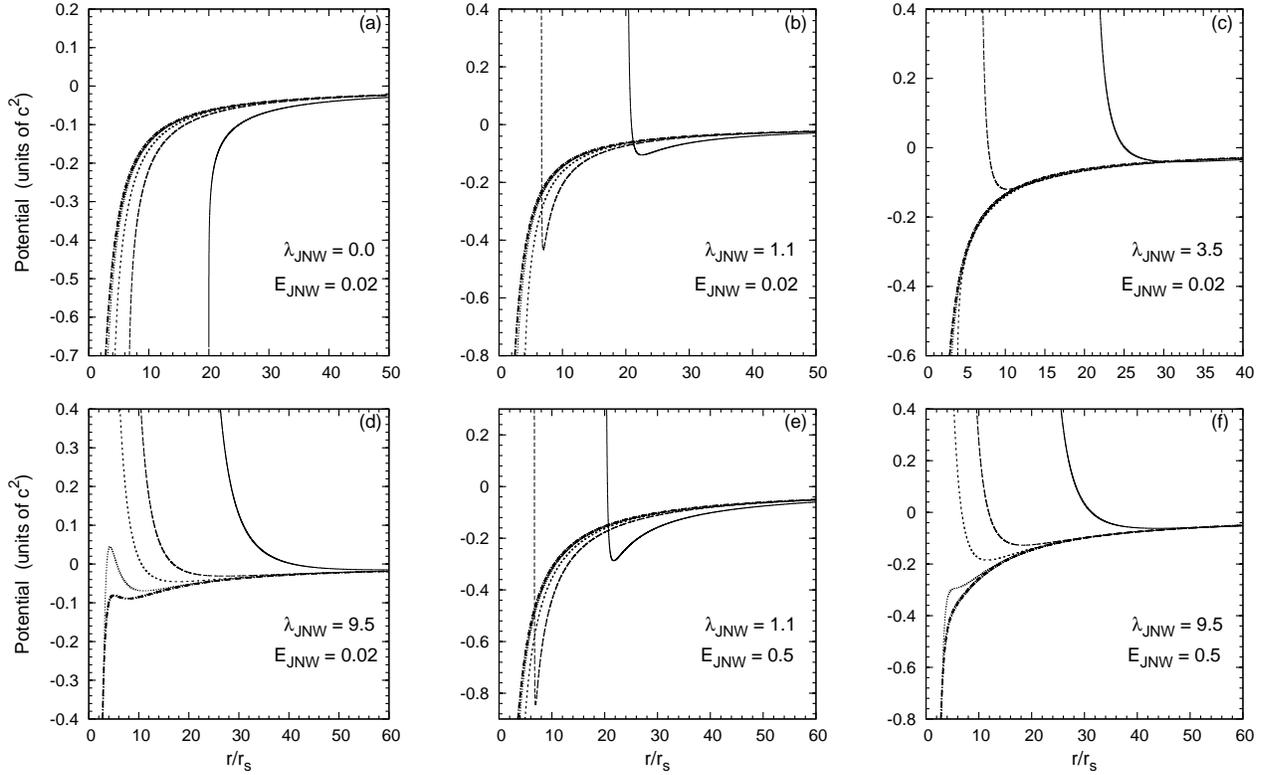}
\end{center}
\caption{Variation of $V_{\rm JNW}$ (Eq. (17)) with radial distance $r$ for different values of $\lambda_{\rm JNW}$. Solid, long-dashed, short-dashed, dotted and long dotted-dashed curves in all the plots are for $\gamma = (0.1, 0.3, 0.5, 0.8, 1)$, respectively. Figures 1a,b,c,d correspond to low energy limit of the test particle motion with $E_{\rm JNW} = 0.02$. Although the nature of profiles are similar for semi-relativistic energy $E_{\rm JNW} = 0.5$, however, for a comparison we show the profiles for $E_{\rm JNW} = 0.5$ corresponding to two different values of $\lambda_{\rm JNW}$. $V_{\rm JNW}$ and $E_{\rm JNW}$ are in units of $c^2$ whereas $\lambda_{JNW}$ is in units of $GM/c$.  }
\label{Fig1} 
\end{figure}

We next obtain the equation of the orbital trajectory using Eqs. (14) and (16), given by 
\begin{eqnarray}
\left(\frac{dr}{d\Omega}\right)^2 = \frac{r^4}{\lambda^2_{\rm JNW}} \left(1-\frac{2r_s}{\gamma r} \right)^{2(1-\gamma)} \, 
\left[2 E_{\rm JNW}  - c^2 \frac{(\gamma r - 2r_s)^{\gamma} - (\gamma r)^{\gamma}}{(\gamma r)^\gamma } - \frac{(\gamma r - 2r_s)^{2\gamma-1}}{(\gamma r)^{2\gamma-1}} \, \frac{\lambda^2_{\rm JNW}}{r^2}  \right] \, , 
\label{18} 
\end{eqnarray}
which exactly matches the corresponding GR expression. Equivalently, the equation of motion in spherical geometry (from the Euler-Lagrange equations), which describes the complete behavior of the test particle dynamics, is given by
\begin{eqnarray}
\ddot r = -c^2 \left(1-\frac{2r_s}{\gamma r} \right)^{3\gamma -1} \frac{r_s}{r^2} 
+ \frac{2 \dot r^{2}}{\left(1-\frac{2r_s}{\gamma r} \right)} \frac{r_s}{r^2} 
+ \left[r -\frac{r_s}{\gamma} (1 + 2 \gamma) \right]  \left( \dot \theta^{2} + \sin^2 \theta \dot \phi^{2} \right)  \, , 
\label{19}
\end{eqnarray}

\begin{eqnarray}
\ddot \phi = -\frac{2\dot r \, \dot \phi}{r} 
\left[ \frac{\gamma r-r_s (1 + 2 \gamma)}{\gamma r-2 r_s } \right] - 
2 \cot\theta \, \dot\phi \, \dot \theta 
\label{20}
\end{eqnarray}
and 
\begin{eqnarray}
\ddot \theta = -\frac{2\dot r \, \dot \theta}{r} 
\left[ \frac{\gamma r-r_s (1 + 2 \gamma)}{\gamma r-2 r_s } \right] 
 + \sin \theta \, \cos \theta \, \dot \phi^2  \, , 
\label{21}
\end{eqnarray}
respectively. The $\ddot \phi$ and $\ddot \theta$ equations are exactly same to those in general relativity, whereas $\ddot r$ in Eq. (20) is similar to that in general relativity in the low energy limit. The corresponding $\ddot r$ equation in general relativity is given by 
\begin{eqnarray}
\ddot r = -\frac{c^6}{\epsilon^2} \, \left(1-\frac{2r_s}{\gamma r} \right)^{3\gamma -1} \frac{r_s}{r^2} 
+ \frac{2 \dot r^{2}}{\left(1-\frac{2r_s}{\gamma r} \right)} \frac{r_s}{r^2} 
+ \left[r -\frac{r_s}{\gamma} (1 + 2 \gamma) \right]  \left( \dot \theta^{2} + \sin^2 \theta \dot \phi^{2} \right)  \, .
\label{22}
\end{eqnarray}

\subsection{Particle dynamics along circular orbit}

We next study the dynamics of the test particle motion in circular orbit in the presence of $V_{\rm JNW}$ and 
compare the behavior of the corresponding test particle dynamics in full general relativity. Using the conditions for the circular orbit $\dot r = 0$ and $\ddot r=0$, we obtain corresponding specific angular momentum $\lambda^C_{\rm JNW}$, 
specific Hamiltonian $E^C_{\rm JNW}$ and specific angular velocity 
$\dot \Omega^{C}_{\rm JNW}$ using $V_{\rm JNW}$, given by 
\begin{eqnarray}
\lambda^{C}_{\rm JNW} = \sqrt{\frac{c^2 r r_s (\gamma r)^\gamma}{(\gamma r - 2r_s)^\gamma - (2\gamma -1) (\gamma r - 2r_s)^{\gamma -1} r_s }  } \, ,
\label{23}
\end{eqnarray}
\begin{eqnarray}
E^{C}_{\rm JNW } = \frac{c^2}{2} \frac{\left(1-\frac{2r_s}{\gamma r}\right)^{1+\gamma} - \left(1-\frac{2r_s}{\gamma r}\right) + \frac{r_s}{\gamma r} 
\left[(1-\gamma)\left(1-\frac{2r_s}{\gamma r}\right)^\gamma + (2\gamma -1)  \right]}{\left(1-\frac{2r_s}{\gamma r}\right) - (2\gamma -1) \frac{r_s}{\gamma r} }
\,
\label{24}
\end{eqnarray}
and 
\begin{eqnarray}
\dot \Omega^{C}_{\rm JNW} = \frac{\left(1-\frac{2r_s}{\gamma r}\right)^{2\gamma -1}}{r^2} \, \sqrt{\frac{c^2 r r_s (\gamma r)^\gamma}{(\gamma r - 2r_s)^\gamma - (2\gamma -1) (\gamma r - 2r_s)^{\gamma -1} r_s }  } \, , 
\label{25}
\end{eqnarray}
respectively. With $\gamma = 1$, the dynamical equations of Schwarzschild geometry are recovered. The corresponding `GR effective potential' and the specific energy $\epsilon$ are given by the relations 
\begin{eqnarray} 
V^{\rm JNW}_{\rm eff} \, (r) = \left(1 - \frac{2 r_s}{\gamma r}\right)^\gamma \, \left[c^2  + \left(1 - \frac{2 r_s}{\gamma r}\right)^{\gamma-1} \, \frac{\lambda^2}{r^2} \right] \,  
\label{26}
\end{eqnarray}
and
\begin{eqnarray}
\frac{\epsilon}{c^2} = \sqrt{  \frac{\left(1 - \frac{2 r_s}{\gamma r}\right)^{2\gamma} - (\gamma -1) \left(1 - \frac{2 r_s}{\gamma r}\right)^{2\gamma-1} \frac{r_s}{\gamma r} }{\left(1 - \frac{2 r_s}{\gamma r}\right)^{\gamma} - (2 \gamma -1) \left(1 - \frac{2 r_s}{\gamma r}\right)^{\gamma-1} \frac{r_s}{\gamma r}  }  } \, , 
\label{27}
\end{eqnarray}
respectively. 
Using Eqs. (26), (27) and with the usual conditions, we obtain specific the angular momentum and equivalent Hamiltonian in general relativity, which are exactly the same as those derived from the potential $V^{\rm JNW}$. The specific angular velocity in general relativity is then given by 
\begin{eqnarray}
\dot \Omega^{C} =  \frac{\left(1-\frac{2r_s}{\gamma r}\right)^{2\gamma -1}}{r^2} \, \sqrt{\frac{c^2 r r_s (\gamma r)^{2\gamma}}{(\gamma r - 2r_s)^{2\gamma} - (\gamma -1) (\gamma r - 2r_s)^{2\gamma -1} r_s }  } \, , 
\label{28}
\end{eqnarray}
where the expression is not exactly equivalent to that obtained using $V^{\rm JNW}$. From Eq. (23), the photon 
orbit or null hypersurface is obtained at $r = r_s (1+2\gamma)/{\gamma}$. With $\gamma = 1$, the usual photon orbit in Schwarzschild geometry is obtained. The timelike circular orbits corresponding to JNW geometry occurs only for $r > r_s (1+2\gamma)/{\gamma}$.

In Fig. 2 we show the radial variation of $\lambda^C_{\rm JNW}$ and $E^C_{\rm JNW}$, obtained using $V^{\rm JNW}$ for various $\gamma$, corresponding to timelike circular geodesics. For $0.5 \, \leq \, \gamma $, the nature of the profiles of both $\lambda^C_{\rm JNW}$ and $E^C_{\rm RN}$ are similar to those around Schwarzschild BH. $\lambda^C_{\rm JNW}$ profiles show that for $0.4472 \, \lsim \, \gamma < 0.5$, apart from a single minima, $\lambda^C_{\rm JNW}$ consists also of another maxima. However, for values of $\gamma < 0.4472$, the profiles of $\lambda^C_{\rm JNW}$ do not show any maxima or minima, implying that the particle in circular trajectory would not have last stable orbit for these values of $\gamma$, corresponding to JNW geometry. The Hamiltonian $E^C_{\rm JNW}$, too, does not attain zero value for $\gamma \, \lsim \, 0.4757$ as may be seen from Fig. 2c, indicating that the circular orbits will always remain bound for $\gamma \, \lsim \, 0.4757$. We also found that specific angular velocity $\dot \Omega^{C}_{\rm JNW}$ obtained using $V^{\rm JNW}$ resembles the GR results with good accuracy for the entire range of $\gamma$, as exemplified in Fig. 3a for two values of $\gamma$. 

\begin{figure}
\begin{center}
\includegraphics[width = 1.0\textwidth,angle=0]{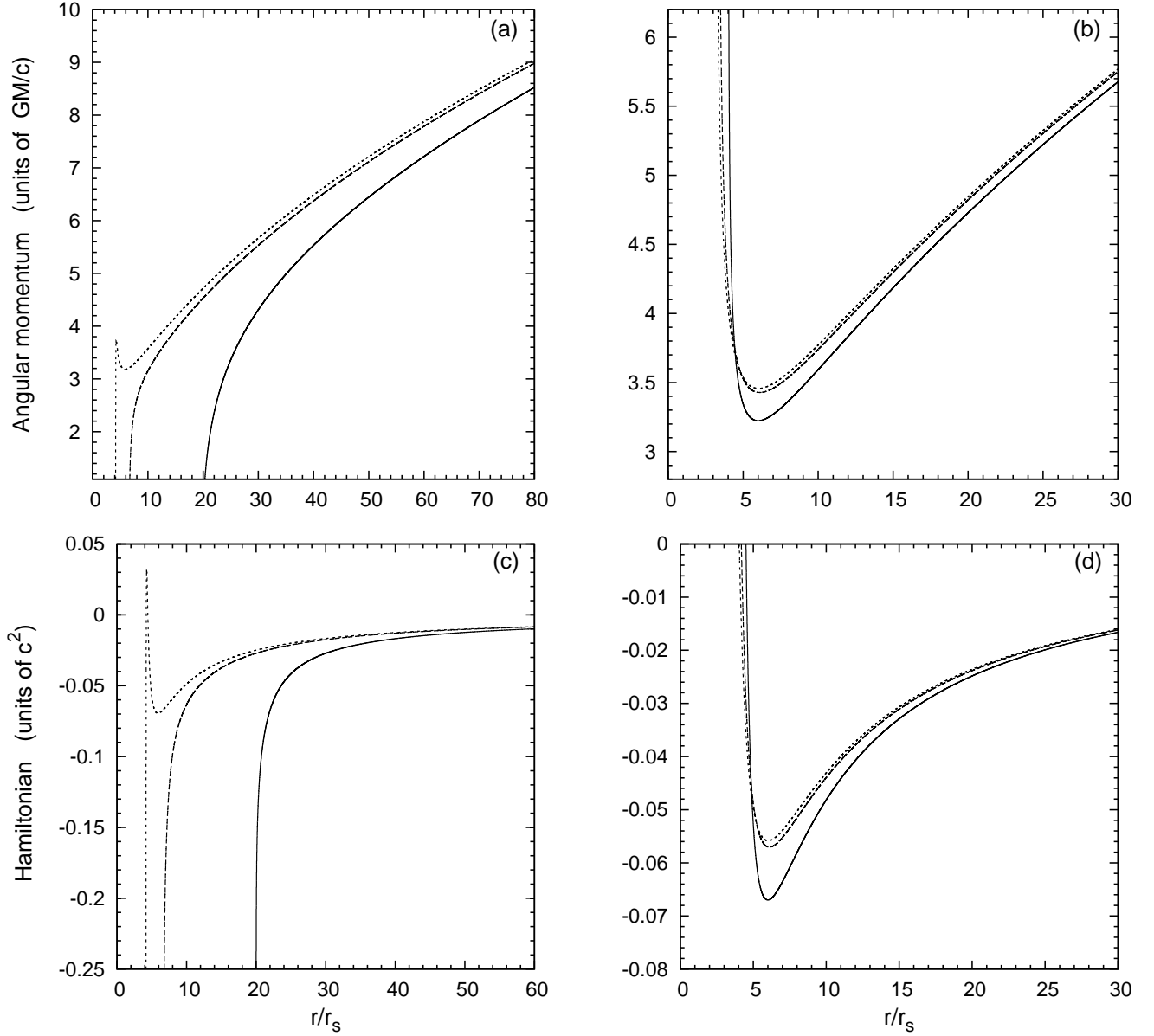}
\end{center}
\caption{Variation of $\lambda_{\rm JNW}$ and $E_{\rm JNW}$ in $r$ for different values of $\gamma$. Solid, long-dashed and short-dashed curves in Fig. 2a are for $\lambda_{\rm JNW}$ corresponding to $\gamma = (0.1, 0.3, 0.48)$, respectively.
Solid, long-dashed and short-dashed curves in Fig. 2b are for $\lambda_{\rm JNW}$ corresponding to $\gamma = (0.5, 0.8, 0.95)$, respectively. Figures 2c,d are similar to those of figures 2a,b, but for $E_{\rm JNW}$. $\lambda_{JNW}$ and $E_{\rm JNW}$ are in units of $GM/c$ and $c^2$, respectively.
 }
\label{Fig2}
\end{figure}

\begin{figure}
\begin{center}
\includegraphics[width = 1.0\textwidth,angle=0]{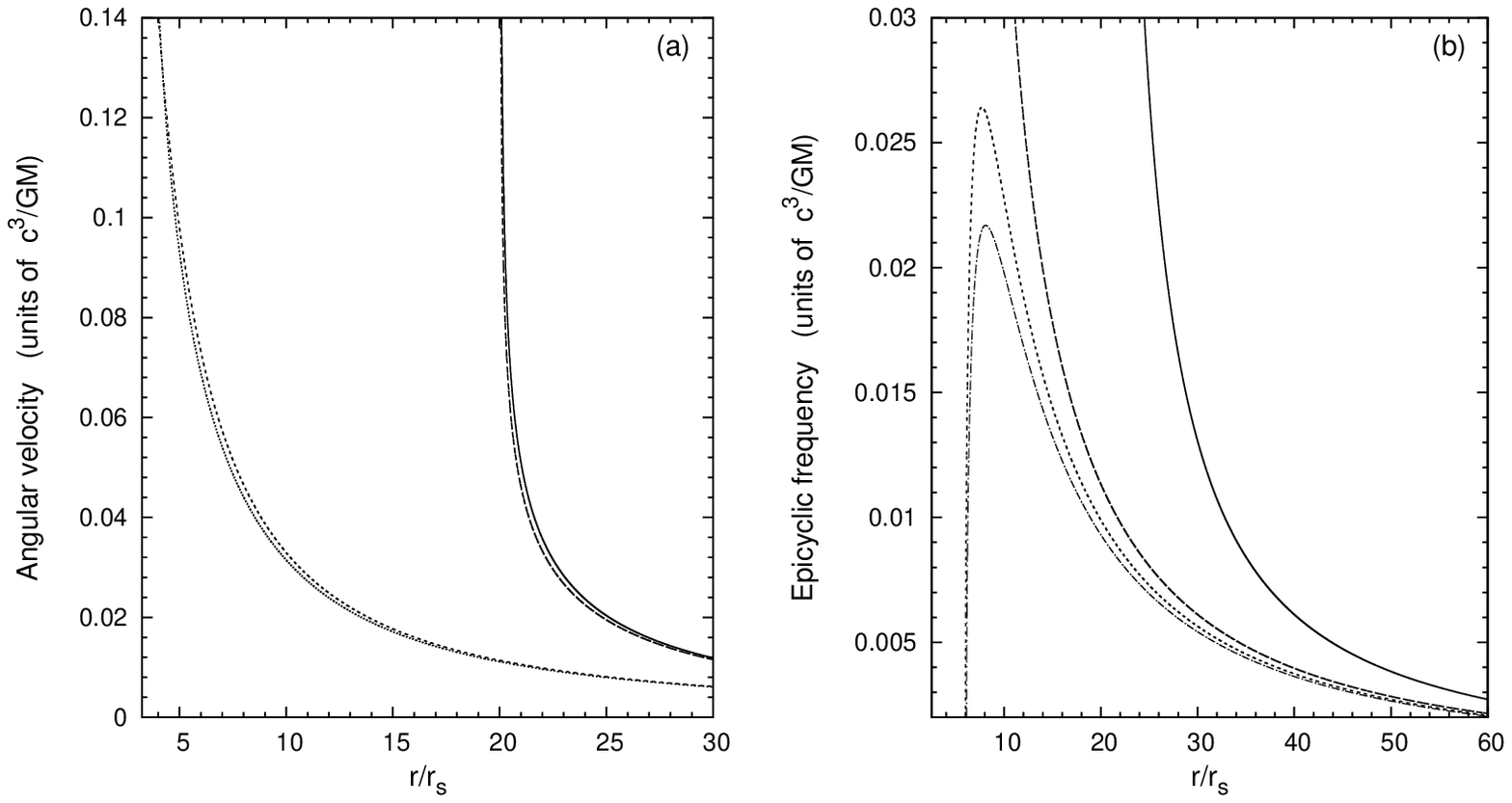}
\end{center}
\caption{Figure 3a shows the comparison of the radial variation of specific angular velocity for two values of $\gamma$ using $V_{\rm JNW}$ and those of corresponding GR results, for timelike circular geodesics. Solid and long-dashed curves correspond to general relativity and $V_{\rm JNW}$, respectively, for a small value of $\gamma ( = 0.1)$. Similarly, Short-dashed and dotted curves correspond to general relativity and $V_{\rm JNW}$, respectively, for a large value of $\gamma ( = 0.8)$. In Fig. 3b we show the variation of radial epicyclic frequency $\kappa$ with $r$ using $V_{\rm JNW}$ for various $\gamma$, for time like circular geodesics. Solid, long-dashed, short-dashed and dotted-dashed curves in Fig. 3b are for $\gamma =(0.1,0.3,0.5,0.8)$, respectively. Specific angular velocity and epicyclic frequency are expressed in units of $c^3/GM$. 
 }
\label{Fig3}
\end{figure}

It is necessary for us to investigate the perturbative effects on the orbital dynamics as the perturbative effects 
can have substantial effect on the accretion flow stability in the vicinity of compact objects. To study the orbital perturbation of the particle orbit around JNW geometry, we compute epicyclic frequency for small perturbation 
of the particle orbit in circular trajectory using $V_{\rm JNW}$, confining to the equatorial plane of test particle motion. It is to be noted that unlike the case in Schwarzschild and Kerr geometries having event horizons [26], till date, there is no exclusive analytical expression of epicyclic frequency in full general relativity for JNW geometry. Owing to naked singularity behavior of JNW spacetime, it would be quite interesting to study the orbital perturbation around these events. The linearized perturbed equations of motion are evaluated using Eqs. (19), (20) and (21), given by
\begin{eqnarray}
\delta {\ddot{r}} = \left[(\gamma r - 2r_s)^{3\gamma -2} \, \left[\gamma r - r_s (1+ 3 \gamma) \right] \frac{2GM}{r^3 (\gamma r)^{3\gamma -1}} + {\dot \phi}^2 \, {\vert}_C  \right]  \, \delta r \nonumber \\ 
+ \, 2 {\dot \phi} \, {\vert}_C \, \left(r - r_s \frac{1+2\gamma}{\gamma} \right) \delta {\dot{\phi}}\, ,
\label{29}
\end{eqnarray}
\begin{eqnarray}
\delta {\ddot{\phi}} = -\frac{2 {\dot \phi} \, {\vert}_C }{r} \, 
\left[ \frac{\gamma r-r_s (1 + 2 \gamma)}{\gamma r-2 r_s } \right] \, \delta{\dot r} 
\label{30}
\end{eqnarray}
and 
\begin{eqnarray}
\delta {\ddot{\theta}} = - {\dot \phi}^2 \, {\vert}_C \, {\delta \theta} \, ,
\label{31}
\end{eqnarray}
respectively. For the particle orbits in equatorial plane, $\dot \Omega^{C}_{\rm JNW}  \equiv {\dot \phi} \, {\vert}_C$. Using the expressions of perturbed quantities for harmonic oscillations given by $\delta r = \delta r_0 \exp^{\imath \kappa t}$ and $\delta \phi = \delta \phi_0 \exp^{\imath \kappa t}$ in Eqs. (29) and (30), where $\kappa$ is the radial epicyclic frequency and $\delta r_0$ and $\delta \phi_0$ are amplitudes (see [8], [9]), we derive the radial epicyclic frequency $\kappa$ after rigorous algebra using JNW analogous potential $V_{\rm JNW}$, which is given by 
\begin{eqnarray}
\kappa = \sqrt{\frac{\gamma^2 r^2 - 2 \gamma r r_s (1 + 3 \gamma) + 2 (1 + \gamma) (1 + 2\gamma) r^2_s }{\gamma r - r_s (1 + 2 \gamma) } \, 
\frac{(\gamma r - 2 r_s)^{3\gamma-2}}{(\gamma r)^{3\gamma-1}} \, \frac{GM}{r^3} } \, .
\label{32}
\end{eqnarray}

The expression in Eq. (32) reduces to that for Schwarzschild geometry with $\gamma = 1$. Although the radial epicyclic frequency has been derived using JNW analogous potential, nonetheless, based on the comparison of the magnitude of $\kappa$ between Schwarzschild analogous potential and corresponding GR result, we too predict here that the radial epicyclic frequency obtained with $V_{\rm JNW}$ would reproduce the GR result with precise/reasonable accuracy, plausibly within a small error margin. For values of $\gamma < 0.4472$, epicyclic frequency monotonically increases in the inward radial 
direction. However, for $\gamma \, \gsim \, 0.4472$, the profiles of epicyclic frequencies resemble the corresponding profile in Schwarzschild geometry (see Fig. 3b).

\subsection{Stability and boundedness of circular orbit} 

We obtain the last stable or marginally stable orbit $\left(r_{\rm ms} \right)$ of the test particle using $V_{\rm JNW}$ with the condition ${d\lambda^{C}_{\rm JNW}}/{dr} = 0$ or an equivalent relation 
\begin{eqnarray}
\gamma^2 r^2 -2 r r_s \gamma (1 + 3 \gamma) + 2r^2_s (1 + 3 \gamma + 2 \gamma^2)
=  0 \, , 
\label{33}
\end{eqnarray}
which is exactly same to that obtained in general relativity for $\gamma=1$. Similarly the marginally bound orbit $\left(r_{\rm mb}\right)$ of the test particle can be obtained using $V_{\rm JNW}$ with the condition $E^C_{\rm JNW} = 0$ or an equivalent relation 
\begin{eqnarray}
\left(\gamma r-2r_s \right)^{1+\gamma} - \left(\gamma r- 2r_s \right) (\gamma r)^\gamma + r_s \left[(1-\gamma) (\gamma r- 2r_s)^\gamma + (2\gamma -1) (\gamma r)^\gamma \right]
=  0 \, , 
\label{34}
\end{eqnarray} 
which exactly matches that in general relativity for $\gamma=1$. With $\gamma = 1$, the familiar circular orbit stability limit and the marginally bound circular orbit for the Schwarzschild metric are recovered. Eq. (33) renders two real and positive roots furnishing two values of $r_{\rm ms}$ for $\gamma < 0.5$. The two real and positive roots corresponding to $r_{\rm ms}$, however, coincide at $\gamma \sim 0.4472$, below which we do not obtain any real and positive value for $r_{\rm ms}$, and consequently there would be no last stable circular orbit for test particle motion (see also Fig. 2). On the other hand, Eq. (34) renders only one real and positive root corresponding to timelike circular geodesics for $\gamma < 0.5$, thus obtaining only one real and positive value for $r_{\rm mb}$. However, the curve for $r_{\rm mb}$ gets truncated at the corresponding value of $\gamma \sim 0.4757$, implying that for values of $\gamma < 0.4757$, the circular orbits will always remain bound. 

Figure 4a shows the variation of $r_{\rm ms}$ and $r_{\rm mb}$ with $\gamma$. In figures 4b,c, we display the variation of 
$E^{C}_{\rm JNW }$ and $\lambda^{C}_{\rm JNW}$ obtained along $r_{\rm ms}$ and $r_{\rm mb}$, with $\gamma$. For 
$0.4757 < \gamma <  0.5$, one of the solutions of Hamiltonian, obtained at $r_{\rm ms}$, gives positive value inferring that even at last stable circular orbit the particle motion may become unbound for such values of $\gamma$. 

\begin{figure}
\begin{center}
\includegraphics[width = 0.95\textwidth,angle=0]{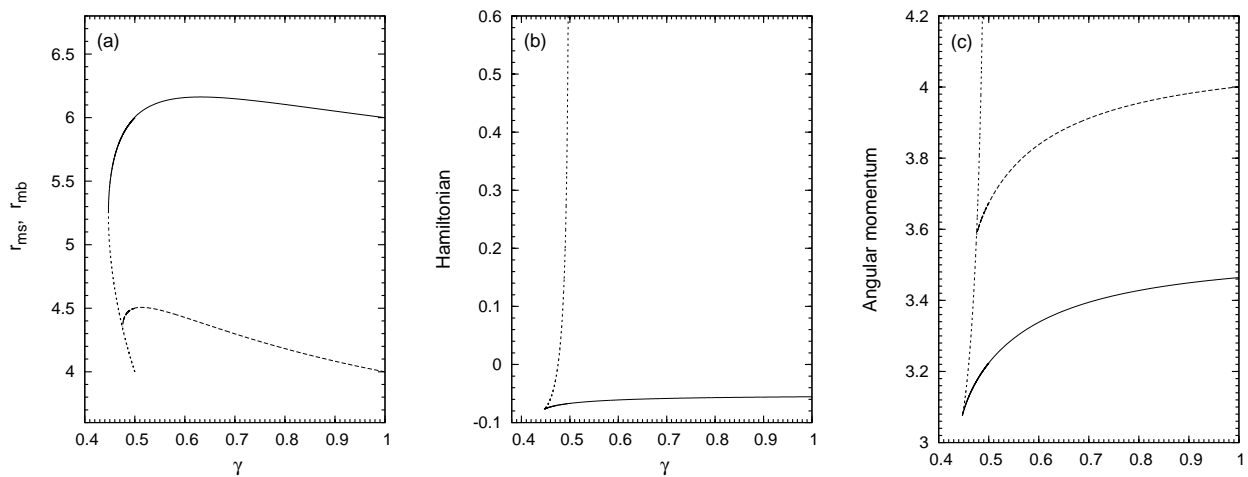}
\end{center}
\caption{Variation of $\left(r_{\rm ms} \right)$ and $\left(r_{\rm mb} \right)$ with $\gamma$ and the nature of dynamical variables along them, corresponding to JNW geometry, obtained using $V_{\rm JNW}$. In Fig. 4a solid curve shows the variation of $r_{\rm ms}$ with $\gamma$ in the range $1 \,\leq \, \gamma \, \leq \, 0.5$, solid and short-dashed curves corresponding to $\gamma \, \leq \, 0.5$ show the variation of $r_{\rm ms}$ for two real and positive roots of timelike circular geodesics. Long-dashed curve shows the variation of $r_{\rm mb}$ with $\gamma$. The curves in Fig. 4b exhibit the variation of $E^C_{\rm JNW}$ along $r_{\rm ms}$ for the curves in Fig. 4a. Similarly the curves in Fig. 4c represent the variation of $\lambda^C_{\rm JNW}$ along $r_{\rm ms}$ and $r_{\rm mb}$ corresponding to curves in Fig. 4a. $r_{\rm ms}$ and $r_{\rm mb}$ are in units of $r_s$, $E^C_{\rm JNW}$ and $\lambda^C_{\rm JNW}$ are in units of $c^2$ and $GM/c$, respectively. 
 }
\label{Fig4}
\end{figure}

Following the similar procedure adopted here, in the next section we analyze the test particle dynamics around RN geometry in the modified Newtonian analogue and compare that with the corresponding GR results.

\section{Orbital dynamics around RN spacetime}

RN metric is a non vacuum static GR counterpart of a Schwarzschild solution in the presence of an electromagnetic field, which describes the exterior gravitational and electromagnetic of an arbitrary-static, oscillating, collapsing or expanding spherically symmetric charged BH of mass $M$ and charge $Q$. The metric function of RN geometry is $f(r)= 1- \frac{2r_s}{r} + \frac{r^2_Q}{r^2}$, where $r^2_Q = {Q^2 G}/{c^4}$ with the arbitrary constant in Eq. (1) $\beta =1$. The 
metric diverges at $r - 2r_s + {r_{Q}^2}/{r} = 0$; for $r_Q \, \leq \, r_s$ it generates two horizons. They are outer (event) horizon and the inner (Cauchy) horizon given by the relation $r_{\pm} = \left(r_s \pm \sqrt{r^2_s - r^2_Q} \right)$, respectively. For $r_Q > r_s$, the above relation generates naked singularities. Outer horizon properties corresponding to 
normal BH with condition $r_Q < r_s$ and extremal BH with condition $r_Q = r_s$. The motion on the other side of the Cauchy horizon is only possible along spacelike geodesics. Using the relation of $f(r)$ in Eq. (11), the three dimensional 
generalized potential in modified Newtonian analogue corresponding to RN spacetime, in the low energy limit, in spherical geometry, is given by  
\begin{eqnarray}
V_{\rm RN} = \left(-\frac{GM}{r} + \frac{c^{2} \, r_{Q}^2}{2r^2} \right) -
\left(\frac{2r_s - \frac{r_{Q}^2}{r}}{r - 2r_s + \frac{r_{Q}^2}{r}} \right)
\left(\frac{r-r_s + \frac{r_{Q}^2}{2r}}{r - 2r_s + \frac{r_{Q}^2}{r}} \, 
{\dot r}^2 + \frac{r^{2}\dot \Omega^{2}}{2} \right) \, ,
\label{35} 
\end{eqnarray}
which we would refer to as RN analogous potential. With $r_Q=0$, the usual Schwarzschild solution and its corresponding properties are recovered. We confine ourselves with the motion along timelike geodesics. 
The Lagrangian per unit mass for this potential is then given by
\begin{eqnarray}
{\cal L}_{\rm RN} = \frac{1}{2} \left[\frac{r^2 \dot r ^{2}}{\left(r - 2r_s + \frac{r_{Q}^2}{r} \right)^2} + \frac{r^3 \dot \Omega^{2}}{\left(r - 2r_s 
+ \frac{r_{Q}^2}{r} \right)} \right]
\, + \, \frac{GM}{r} - \frac{c^2 r_{Q}^{2}}{2r^2}  \, , 
\label{36}
\end{eqnarray}
where $\dot \Omega$ is described in \S 3. As usual, all the relevant dynamical quantities and the geodesic equations of motion for RN analogous potential $V_{\rm RN}$ can be easily evaluated/computed using the Lagrangian described in 
Eq. (36), following \S 3. We do not show the explicit expressions here. The corresponding dynamical expressions could also be computed from the generic expressions given in [23]. Nonetheless, as mentioned earlier that the generic formulation in that paper to derive Newtonian analogous potentials and their dynamical properties corresponding to spherically symmetric metric could not be extended to spacetime geometries coupled to scalar fields describing naked singularities. This is in contrast to the formulation presented in this work, which is the most generalized generic formulation of the potential analogue of any static spherically symmetric GR geometry having vacuum or non vacuum solutions even with any arbitrary scalar field describing event horizons and/or naked singularities. 

In terms of conserved Hamiltonian $E_{\rm RN}$ and angular momentum $\lambda_{\rm RN}$, the RN analogous potential in Eq. (35) can be written as   
\begin{eqnarray}
V_{\rm RN}  = \left(-\frac{GM}{r} + \frac{c^{2} \, r_{Q}^2}{2r^2} \right) 
- \left(2r_s - \frac{r_{Q}^2}{r}\right) \left[\left(r-2r_s 
+ \frac{r_{Q}^2}{r}\right) \frac{\lambda^2_{\rm RN}}{r^4} \left(\frac{1}{2} - 
\frac{r-r_s + \frac{r_{Q}^2}{2r}}{r} \right)\right] \nonumber \\
- \left(2r_s - \frac{r_{Q}^2}{r} \right)\left[\frac{1}{r^2} \left(r - r_s + \frac{r_{Q}^2}{2r} \right)  \left(2 E_{\rm RN}  + \frac{2GM}{r} - \frac{c^{2} \, r^2_Q}{r^2} \right)  \right] \, .
\label{37}
\end{eqnarray}
In Fig. 5, we depict the radial profiles for $V_{\rm RN}$ as given in Eq. (37) for $r_Q < r_s$, $r_Q = r_s$, and $r_Q > r_s$ for different values of angular momentum $\lambda_{\rm RN}$. The profiles for $V_{\rm RN}$ show sharp contrast in its behavior for BH solutions with those of naked singularities.

\begin{figure}
\begin{center}
\includegraphics[width = 1.0\textwidth,angle=0]{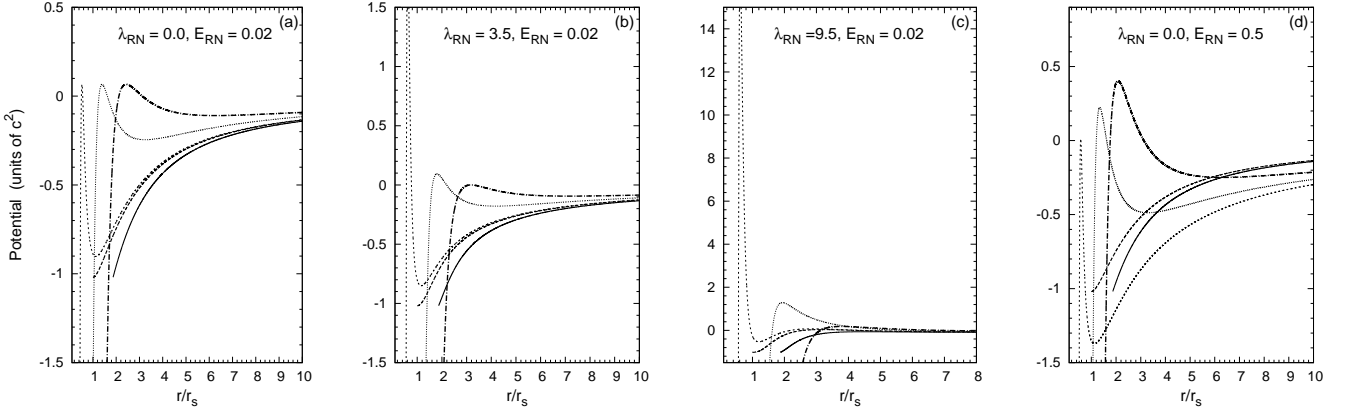}
\end{center}
\caption{Variation of $V_{\rm RN}$ with radial distance $r$ for different $\lambda_{\rm RN}$. Solid, long-dashed, short-dashed, dotted and long dotted-dashed curves in all the plots are for $r_Q/r_s = (0.5, 1, 1.061, 1.8, 2.5)$ respectively. Figures 5a,b,c correspond to low energy limit of the test particle motion with $E_{\rm RN} = 0.02$. Although the nature of profiles are similar for semi-relativistic energy $E_{\rm RN} = 0.5$, however for a comparison we show the profile for $E_{\rm RN} = 0.5$ corresponding to $\lambda_{\rm RN} = 0$. $V_{\rm RN}$ and $E_{\rm RN}$ are in units of $c^2$ whereas $\lambda_{RN}$ is in units of $GM/c$.
 }
\label{Fig5}
\end{figure}

\subsection{Particle dynamics along circular orbit}

We next study the dynamics of the test particle motion in circular orbit in the presence of $V_{\rm RN}$ and 
compare the behavior of the corresponding test particle dynamics in full general relativity, following the case of JNW geometry as in \S 3. The timelike circular orbits corresponding to RN geometry occurs only for  
$r  > \frac{1}{2} \left(3r_s + \sqrt{9r^2_s - 8 r^2_Q}\right)$ and/or 
$r  >  {r^2_Q}/{r_s}$. $r = \frac{1}{2} \left(3r_s + \sqrt{9r^2_s - 8 r^2_Q}\right)$ is the null hypersurface or photon orbit. Specific angular momentum $\lambda^C_{\rm RN}$ and specific Hamiltonian $E^C_{\rm RN}$ corresponding to timelike circular geodesics are obtained using $V_{\rm RN}$ which exactly match the corresponding GR results.    

In Fig. 6 we show the radial variation of $\lambda^C_{\rm RN}$, and $E^C_{\rm RN}$ for different $\gamma$. For $r_Q \, \leq \, \sqrt{9/8} \, r_s$, the nature of the profiles of both $\lambda^C_{\rm RN}$ and $E^C_{\rm RN}$ are similar to those around Schwarzschild BH. However for $r_Q > \sqrt{9/8} \, r_s$ (describing naked singularities), the $\lambda^C_{\rm RN}$ profiles apart from a single minima, consist of another maxima till $r_Q \sim 1.118 \, r_s$; at that value of $r_Q$, both the minima and the corresponding maxima coincide. Beyond which, the profiles of $\lambda^C_{\rm RN}$ do not show any maxima or minima, implying that beyond this value the particle in circular trajectory do not have last stable orbit corresponding to RN geometry. In a similar fashion, beyond $r_Q \sim 1.10887 \, r_s$, $E^C_{\rm RN}$ attains zero value indicating that the circular orbits will always remain bound for $r_Q > 1.10887 \, r_s$ (figures 6d,f).

\begin{figure}
\begin{center}
\includegraphics[width = 1.0\textwidth,angle=0]{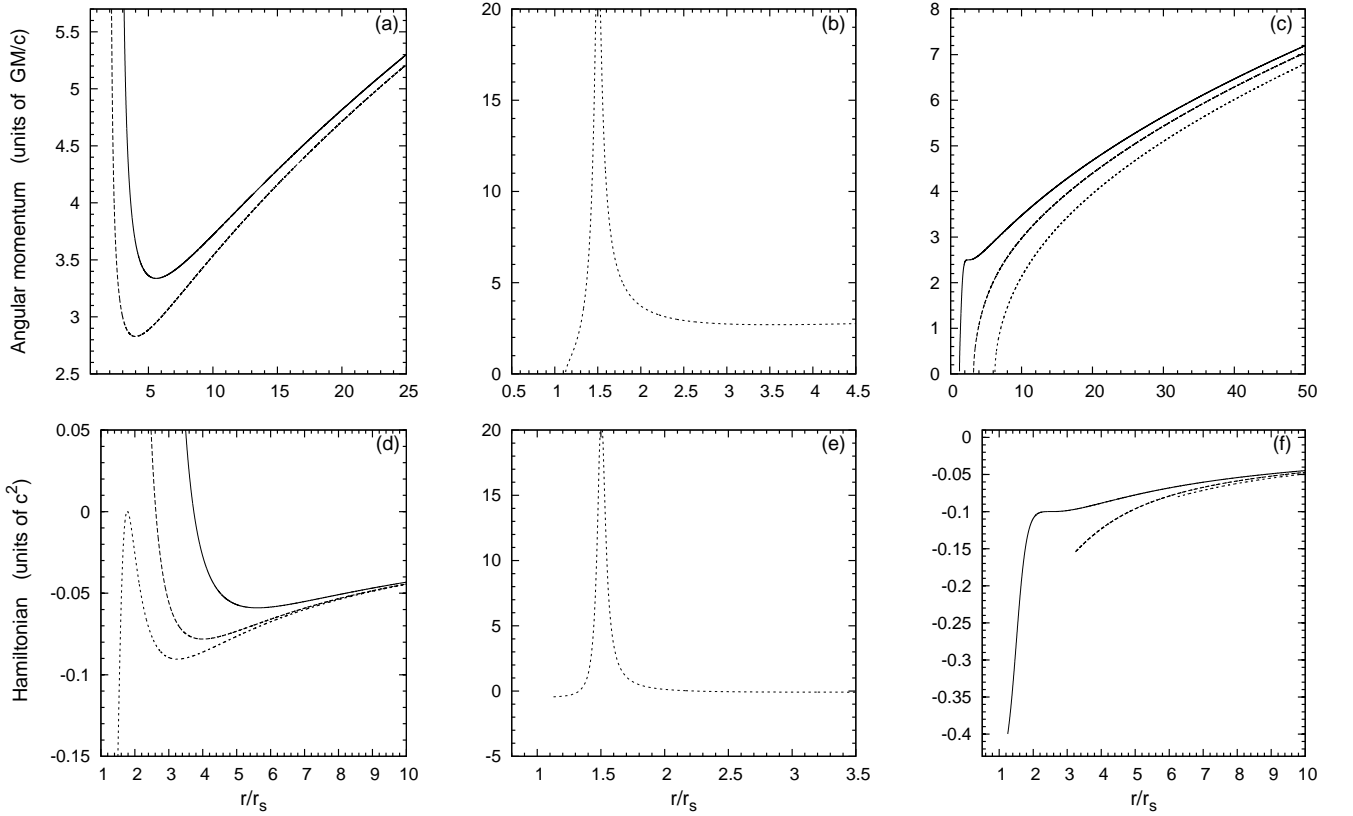}
\end{center}
\caption{Radial profile of $\lambda_{\rm RN}$ and $E_{\rm RN}$ obtained from $V_{\rm RN}$ for different $r_Q$. Solid and long-dashed curves in Fig. 6a are for $\lambda_{\rm RN}$ corresponding to $r_Q/r_s = (0.5, 1)$ respectively. 
Figure 6b is for $\lambda_{\rm RN}$ corresponding to $r_Q/r_s = 1.061$. Solid, long-dashed and shot-dashed curves in Fig. 6c are for $\lambda_{\rm RN}$ corresponding to $r_Q/r_s = (1.118, 1.8, 2.5)$ respectively. Solid, long-dashed and shot-dashed curves in Fig. 6d are for $E_{\rm RN}$ corresponding to $r_Q/r_s = (0.5, 1, 1.0887)$ respectively. Figures 6e,f are similar to that of figures 6b,c, but generated for $E_{\rm RN}$.  
 }
\label{Fig6}
\end{figure}

Fig. 7a shows that for $r_Q \, \leq \, \sqrt{9/8} \, r_s$, the nature of angular velocity profiles are similar to those around Schwarzschild BH, and $\dot \Omega^{C}_{\rm RN}$ (obtained using $V^{\rm JN}$) resembles the corresponding GR counterparts well. For $r_Q > \sqrt{9/8} \, r_s$, the angular velocity profiles show different behavior and the percentage deviation between $\dot \Omega^{C}_{\rm RN}$ and the corresponding GR value becoming large 
($\sim 17 \%$ for $r_Q \sim 1.118 r_s$), however with the further increase in the value of $r_Q$ the error margin between them diminishes significantly ($\sim 9 \%$, for $r_Q \sim 2.5 r_s$) (Fig. 7b). 

\begin{figure}
\begin{center}
\includegraphics[width = 1.0\textwidth,angle=0]{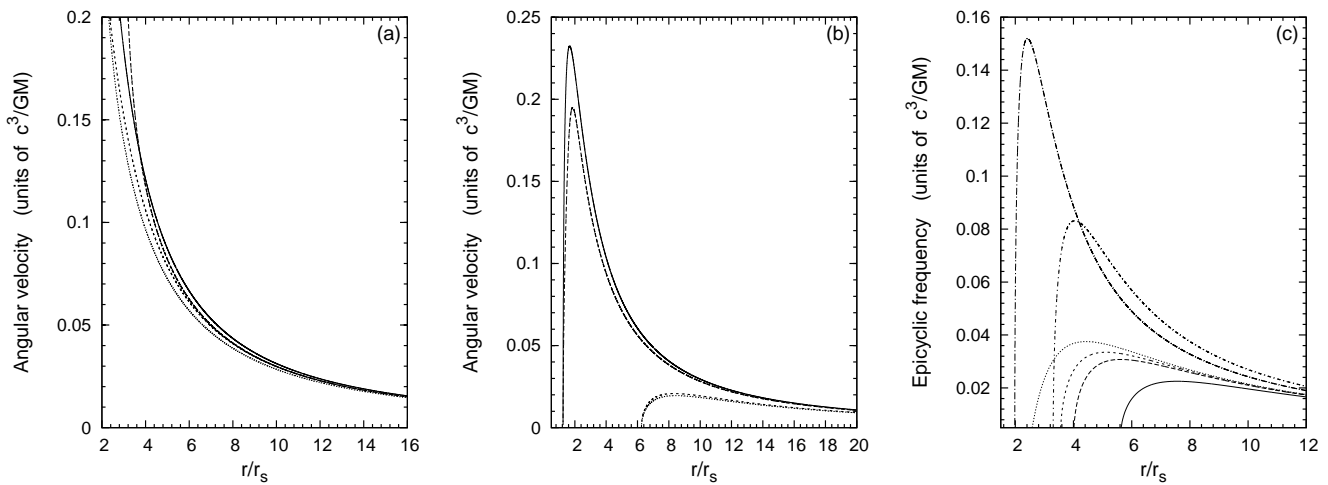}
\end{center}
\caption{Figures 7a,b show the comparison of the radial variation of specific angular velocity for different $r_Q$ (obtained using $V_{\rm RN}$) with the corresponding GR results, for timelike circular geodesics. In Fig. 7a, solid and long-dashed curves correspond to general relativity and $V_{\rm RN}$ respectively, for $r_Q = 0.5 \, r_s$; short-dashed and dotted curves correspond to general relativity and $V_{\rm RN}$ respectively, for $r_Q = \sqrt{9/8} \, r_s$. In Fig. 7b, solid and long-dashed curves correspond to general relativity and $V_{\rm RN}$ respectively, for $r_Q = 1.118 \, r_s$; short-dashed and dotted curves correspond to general relativity and $V_{\rm RN}$, respectively, for $r_Q = 2.5 \, r_s$. In Fig. 7c, we show the variation of radial epicyclic frequency $\kappa$ with $r$ using $V_{\rm RN}$ for various $r_Q$. Solid, long-dashed, short-dashed, dotted, short dotted-dashed, and long dotted-dashed curves in Fig. 7c are for $r_Q/r_s =(0.5, 1, 1.061, 1.118, 1.8, 2.5)$ respectively. Specific angular velocity and epicyclic frequency are expressed in units of $c^3/GM$ 
 }
\label{Fig7}
\end{figure}

To study the orbital perturbation of the particle orbit around RN geometry we compute the radial epicyclic frequency using RN analogous potential $V^{\rm RN}$, restricting ourselves in the equatorial plane of test particle orbit in the circular trajectory. Owing to RN spacetime having three different solutions: ordinary BH solution ($r_Q < r_s$),  
extremal BH ($r_Q = r_s$) and naked singularity ($r_Q > r_s$), it would be quite interesting to study the orbital perturbation around these three respective events. Following the similar procedure adopted in the case of JNW geometry we derive the radial epicyclic frequency $\kappa$, which is given by  
\begin{eqnarray}
\kappa = \left(\frac{r-2r_s + \frac{r^2_Q}{r}}{r - 3r_s + \frac{2 r^2_Q}{r}} \right)^{1/2} \, \sqrt{ \left[ \frac{GM}{r^5} (r-6r_s)(r-2r_s) 
+ \frac{r^2_Q \, c^2}{r^5} \left[ 2r_s \left(5 - 12 \frac{r_s}{r} \right) 
- \frac{r^2_Q}{r} \left(4 + 4 \frac{r^2_Q}{r^2} - 17 \frac{r_s}{r} \right) \right] \right] } \, .
\label{38}
\end{eqnarray}
The expression in Eq. (38) reduces to that in Schwarzschild geometry with $r_Q=0$. As argued in the case of JNW geometry, it is expected that the radial epicyclic frequency computed using $V_{\rm RN}$ would also reproduce the GR result with precise/reasonable accuracy. In Fig. 7c, we show the variation of radial epicyclic frequency $\kappa$ with $r$ for various values of $r_Q$ corresponding to both BH solutions and naked singularities, for time like circular geodesics. 

\subsection{Stability and boundedness of circular orbit}

As usual, the last stable circular orbit $\left(r_{\rm ms} \right)$ and the marginally bound circular orbit $\left(r_{\rm mb} \right)$ of the test particle motion using RN analogous potential $V_{\rm RN}$ can be obtained from the relations 
${d\lambda^{C}_{\rm RN}}/{dr} = 0$ and $E^C_{\rm JNW} = 0$, respectively, which 
are exactly same to the corresponding GR expressions.  

With $r_Q = 0$, the familiar circular orbit stability limit and the marginally bound circular orbit for the Schwarzschild metric are recovered. For 
$r_Q > \sqrt{9/8} \, r_s$, both Eqs. (39) and (40) render two real and positive roots corresponding to timelike circular geodesics. Thus we obtain two values of $r_{\rm ms}$ and $r_{\rm mb}$ for $r_Q > \sqrt{9/8} \, r_s$. The two real and positive roots corresponding to $r_{\rm ms}$, however, coincide at $r_Q \sim 1.118 \, r_s$, beyond which we do not obtain any real and positive value for $r_{\rm ms}$. This signifies that for values of $r_Q > 1.118 \, r_s$, there would be no last stable circular orbit for particle motion. Similarly for $r_{\rm mb}$, the two real and positive roots 
coincide at $r_Q \sim  1.10887 \, r_s$, beyond which we do not obtain any real and positive value for $r_{\rm mb}$, implying that for $r_Q > 1.10877 \, r_s$, the circular orbits will always remain bound. 
Figure 8a shows the variation of $r_{\rm ms}$ and $r_{\rm mb}$ with corresponding values of $r_Q$. In figures 8b,c,d, we display the variation of $E^{C}_{\rm RN }$ and $\lambda^{C}_{\rm RN}$ obtained along $r_{\rm ms}$ and $r_{\rm mb}$, with the corresponding values of $r_Q$. For $\sqrt{9/8} \, r_s < r_Q <  1.10887 \, r_s$, one of the solution for Hamiltonian obtained at $r_{\rm ms}$ gives positive value (Fig. 8b), inferring that even at last stable circular orbit the particle motion may become unbound for those values of $r_Q$. 

\begin{figure}
\begin{center}
\includegraphics[width = 1.0\textwidth,angle=0]{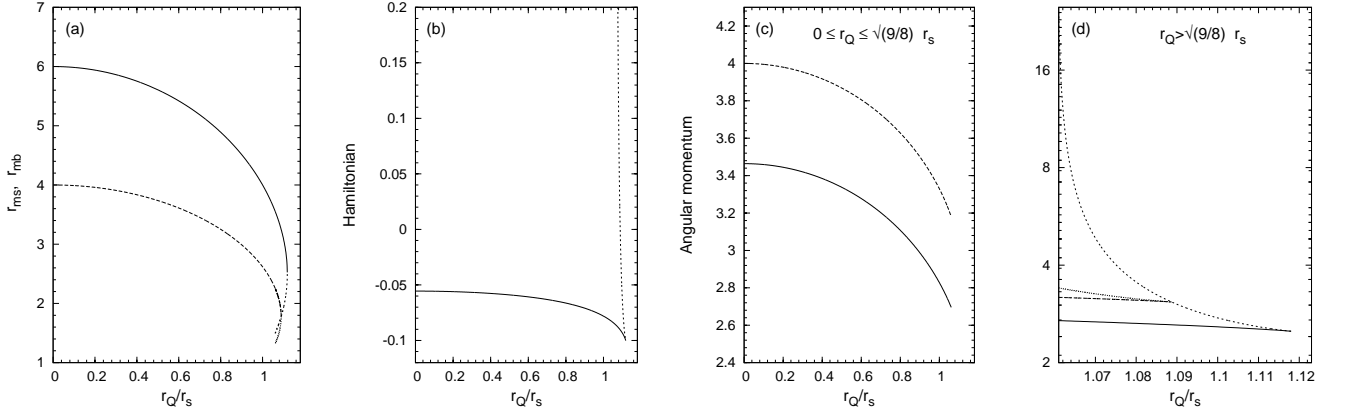}
\end{center}
\caption{Variation of $\left(r_{\rm ms} \right)$ and $\left(r_{\rm mb} \right)$ with $r_Q$ and the nature of dynamical variables along them for RN geometry (using $V_{\rm RN}$). In Fig. 8a solid curve shows the variation of 
$r_{\rm ms}$ with $r_Q/r_s$ in the range $0 \,\leq \, r_Q \, \leq \, \sqrt{9/8} \, r_s$, solid and short-dashed curves  show the variation of $r_{\rm ms}$  with $r_Q/r_s$ in the range $r_Q > \sqrt{9/8} \, r_s$ for two real and positive roots of timelike circular geodesics, long-dashed curve shows the variation of $r_{\rm mb}$ with $r_Q/r_s$ over $0 \,\leq \, r_Q \, \leq \, \sqrt{9/8} \, r_s$, long-dashed and dotted curves show the variation of $r_{\rm mb}$ with $r_Q/r_s$ over $r_Q > \sqrt{9/8} \, r_s$ for two real and positive roots for timelike circular geodesics. The curves in Fig. 8b represent the variation of $E^C_{\rm RN}$ along $r_{\rm ms}$ corresponding to the curves in Fig. 8a. Similarly the curves in figures 8c,d represent the variation of $\lambda_{\rm RN}$ along $r_{\rm ms}$ and $r_{\rm mb}$ corresponding to the curves in Fig. 8a. $r_{\rm ms}$ and $r_{\rm mb}$ are in units of $r_s$. $E^C_{\rm RN}$ and $\lambda^C_{\rm RN}$ are in units of $c^2$ and $GM/c$, respectively. 
 }
\label{Fig8}
\end{figure}

\section{Orbital trajectories}

The dynamics of orbital trajectories for JNW metric in the modified Newtonian analogue can be obtained from the relation for ${d\Omega}/{dr}$ as described in Eq. (18), which is identical to that of general relativity. This implies that $V_{\rm JNW}$ will exactly reproduce the general relativistic trajectories of particle orbits. Similarly, $V_{\rm RN}$ would also exactly replicate the general relativistic trajectories of particle orbits in RN geometry. Consequently, GR apsidal precession and the gravitational lensing would be accurately reproduced by both JNW and RN analogous potential, which are among the few observational tests of general relativity. Following [8] and [9], we show the trajectory profiles of the test particle orbit using $V_{\rm JNW}$ and $V_{\rm RN}$ in the equatorial plane ($x-y$ plane), obtained from the equations of motion. 

Figures 9 and 10 show the elliptic like trajectories of the particle orbits using $V_{\rm JNW}$ and $V_{\rm RN}$ corresponding to JNW and RN geometries for different $\gamma$ and $r_Q$ respectively. Those in Newtonian and Schwarzschild cases are also given in figures 9 and 10 for a comparison. The elliptic like trajectory profiles show clear precession of orbits for all values of $\gamma$ and $r_Q$. For both these JNW and RN geometries, the test particle starts tangentially from a fixed apoapsis $r_a$ with a fixed initial velocity $v_{\rm in} = 0.092 \, c$, for all corresponding values of $\gamma$ and $r_Q$. However to compute the apsidal precession, instead of fixing $v_{\rm in}$, we fix an unique value of eccentricity $e$ for elliptical orbits, for all values of $\gamma$ and $r_Q$. 

\begin{figure}
\begin{center}
\includegraphics[width = 1.0\textwidth,angle=0]{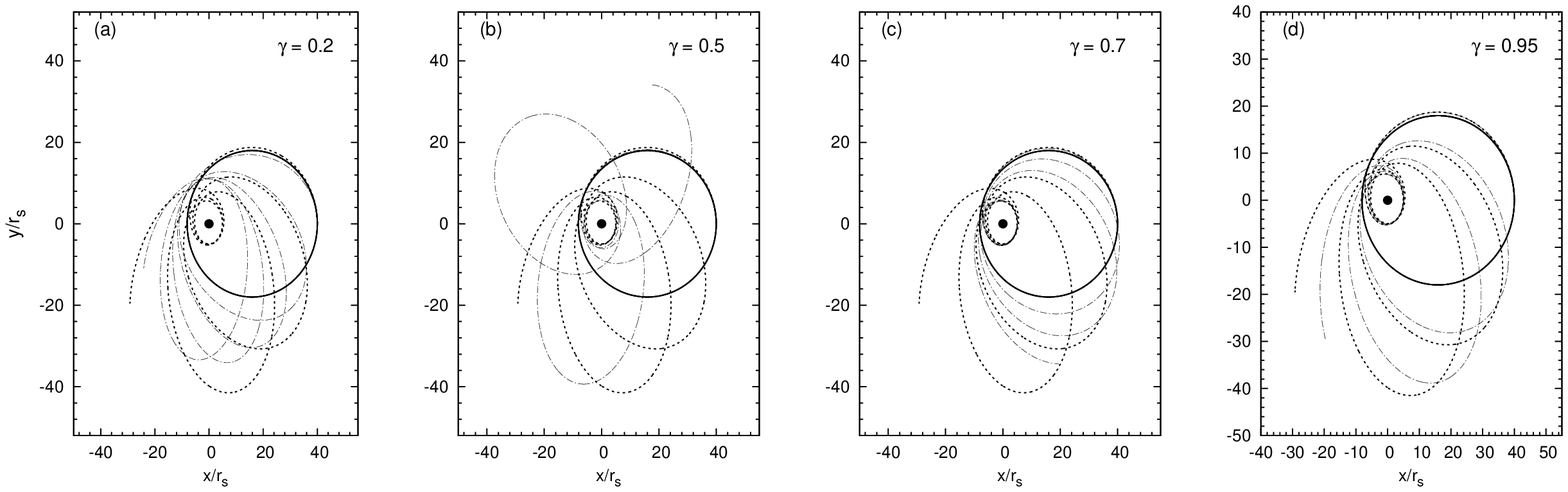}
\end{center}
\caption{ Comparison of elliptic like trajectory of particle orbit in equatorial plane in JNW spacetime with those in Schwarzschild and Newtonian cases projected in x-y plane. Solid and short-dashed lines in all the figures are for Newtonian and Schwarzschild cases respectively. Long dotted-dashed curve in figures 9a,b,c,d are for $\gamma = 0.2, 0.5, 0.7, 0.95$ respectively (using $V_{\rm JNW}$). The particle starts from apogee with $r_a = 40 r_s$ with $v_x = 0.0$ and $v_y \equiv v_{\rm in} = 0.092$. The velocities are expressed in units of $c$. We have restricted down to $\gamma = 0.2$, as for $\gamma < 0.2$, no proper well defined elliptic like orbits are produced with the preferred orbital parameters chosen here.   
 }
\label{Fig9}
\end{figure}

\begin{figure}
\centering
\includegraphics[width=1.0\columnwidth]{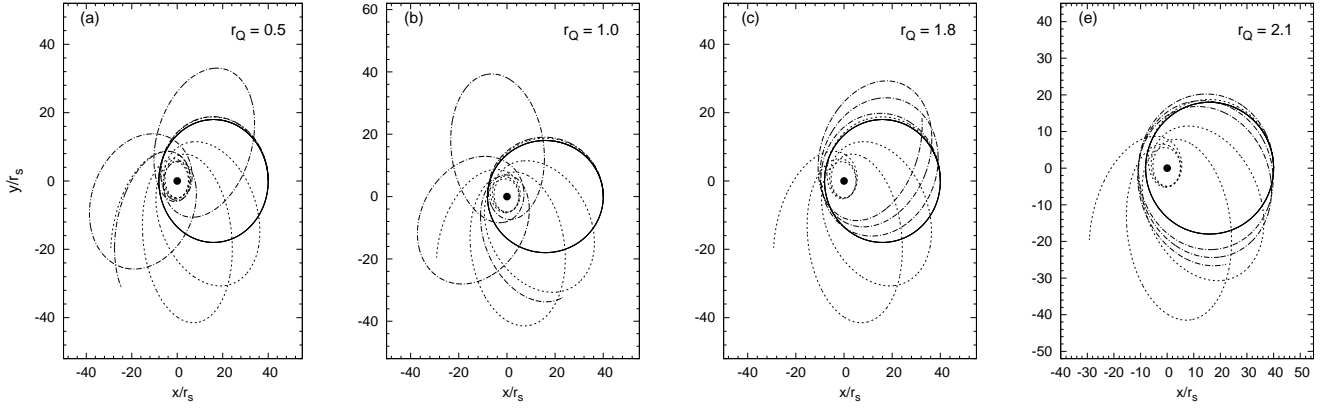}
\caption{ Comparison of elliptic like trajectory of particle orbit in equatorial plane in RN spacetime with those in Schwarzschild and Newtonian cases projected in x-y plane. Solid and short-dashed lines in all the figures are for Newtonian and Schwarzschild case, respectively. Long dotted-dashed curve in figures 10a,b,c,d are for $r_Q/r_s = 0.5, 1, 1.8, 2.1$, respectively, using $V_{\rm RN}$. Other parameters are identical to that of Fig. 9.  }
\label{Fig10}
\end{figure}

Next we compute the apsidal precession or the perihelion advancement $\Psi$ of elliptical orbits for both JNW and RN geometries using the corresponding expressions for ${d\Omega}/{dr}$, given by,  
\begin{eqnarray}
\Psi = \Pi - \pi \equiv \int_{r_p}^{r_a} \frac{d\Omega}{dr} \, dr - \pi  \, ,
\label{39}
\end{eqnarray}
where $\Pi$  is the usual half orbital period of the test particle. $r_p$ and $r_a$ are periapsis and apoapsis of the orbit, respectively. Alternatively, $\Psi$ can be computed directly from the elliptical trajectory profiles. In Fig. 11, we show the variation of $\Psi$ with $\gamma$ computed using $V_{\rm JNW}$, for two scenarios; in one scenario keeping $r_p$ fixed $r_a$ is allowed to varry, whereas in other case $r_a$ is kept fixed while $r_p$ is allowed to varry. For all the cases, the profiles show that with the decrease in $\gamma$, i.e., as one departs from the Schwarzschild BH solution, the magnitude of $\Psi$ continuously increases till $\gamma \sim 0.45$, beyond this value of $\gamma$, the particle trajectory does not produce well defined orbits. In Fig. 12, we show the variation of $\Psi$ with $r_Q$ computed using $V_{\rm RN}$, corresponding to the identical scenarios as investigated for JNW geometry. For all the cases the profiles show that with the increase in $r_Q$, i.e., as one departs from the Schwarzschild BH solution, the magnitude of $\Psi$ decreases till $\Psi$ attains a zero value (like that of the Newtonian case) corresponding to a particular value of $r_Q$ (say $r_Q \vert_N$) describing naked singularity which depends on orbital parameters $r_a$ and $r_p$. However, beyond this value of $r_Q \vert_N$, the magnitude of $\Psi$ again increases, with the particle orbit showing retrograde precession. 

This aspect of retrograde precession for particle orbit around RN geometry for naked singularities can also be found from 
Fig. 10d. Interestingly it is found from Fig. 12 that at $r_Q \sim 1.68 \, r_s$, the magnitude of $\Psi$ corresponding to different values of orbital parameters $r_a$ and $r_p$ in RN geometry becomes almost identical, i.e., for a particular value of $r_Q$ ($r_Q \sim 1.68 \, r_s$), the value of $\Psi$ becomes independent of orbital parameters for elliptical orbits in RN geometry. 

\begin{figure}
\begin{center}
\includegraphics[width = 0.5\textwidth,angle=0]{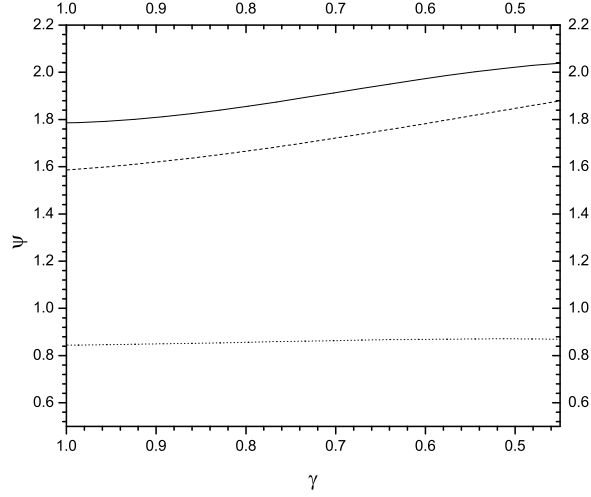}
\end{center}
\caption{Variation of apsidal precession $\Psi$ with $\gamma$ for JNW geometry. Solid and short-dashed curves correspond to $r_p = 6 \,r_s$, with the particle starting from apogee at $r_a = (40, 80) \, r_s$, respectively. Dotted curve correspond to $r_p = 10 \,r_s$, with the particle starting from apogee at $r_a = 40 \, r_s$. $\Psi$ is expressed in radian. 
 }
\label{Fig12}
\end{figure}

\begin{figure}
\begin{center}
\includegraphics[width = 0.5\textwidth,angle=0]{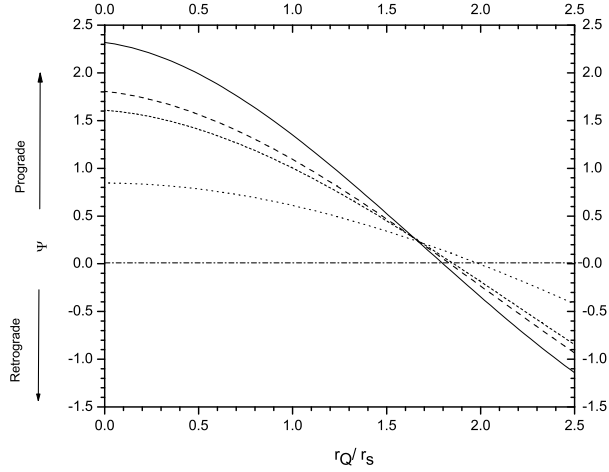}
\end{center}

\caption{Variation of apsidal precession $\Psi$ with $r_Q$ corresponding to RN geometry. Solid, long-dashed, short-dashed curves correspond to $r_p = 6 \,r_s$, with the particle starting from apogee at $r_a = (20, 40, 80) \, r_s$, respectively. The corresponding values of $r_Q \vert_N \sim (1.803, 1.84, 1.853) \, r_s$, respectively. Dotted curve correspond to $r_p = 10 \,r_s$, with the particle starting from apogee at $r_a = 40 \, r_s$. The corresponding value of 
$r_Q \vert_N \sim 1.99 \, r_s $. $\Psi$ is expressed in radian. Dotted-dashed curve represents Newtonian case.
 }
\label{Fig12}
\end{figure}

\begin{figure}
\begin{center}
\includegraphics[width = 1.0\textwidth,angle=0]{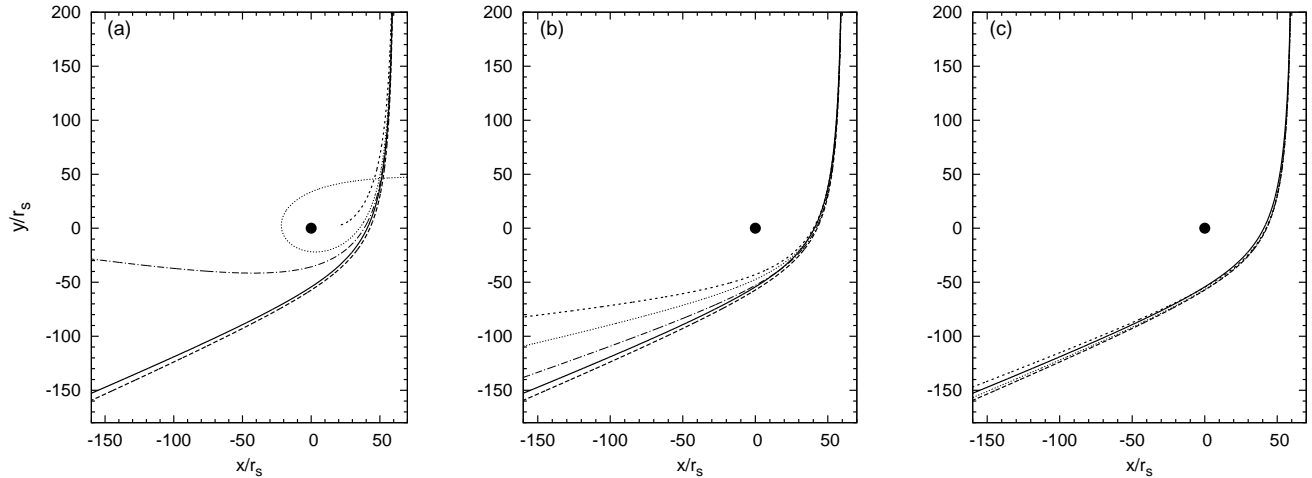}
\end{center}
\caption{Comparison of parabolic like trajectory of particle orbit in equatorial plane in JNW spacetime with that in Schwarzschild and Newtonian case using $V_{\rm JNW}$ with $v_x = 0.0$ and $v_y  =  - 0.1732$. The particle starts from $S_{(x,y)} = (60, 400) \, r_s$. Solid and long-dashed curves in all the figures denote Newtonian and Schwarzschild cases, respectively. Short-dashed, dotted and long dotted-dashed curves in figures 13a,b correspond to $\gamma = (0.1, 0.2, 0.3)$ and $\gamma = (0.4, 0.5, 0.7)$, respectively. Short dashed and dotted curves in Fig. 13c correspond to $\gamma = (0.8, 0.95)$, respectively. The velocities are expressed in units of $c$.
 }
\label{Fig13}
\end{figure}

\begin{figure}
\begin{center}
\includegraphics[width = 1.0\textwidth,angle=0]{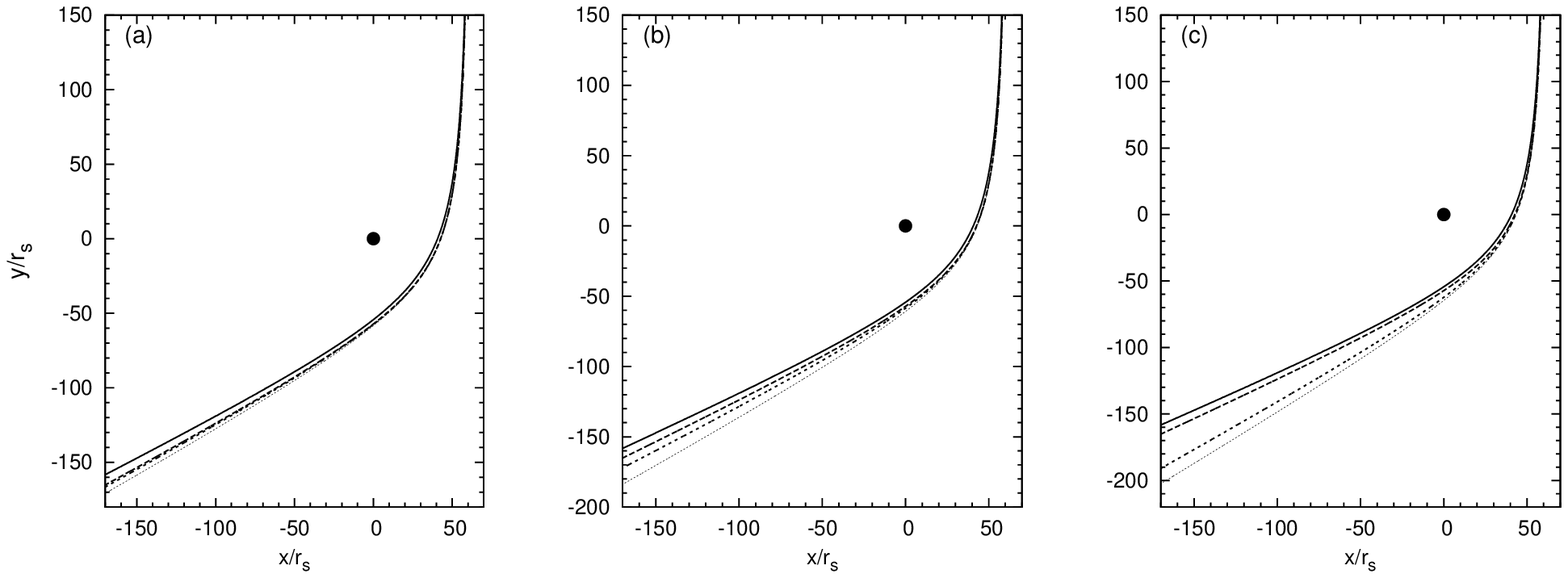}
\end{center}
\caption{Comparison of parabolic like trajectory of particle orbit in equatorial plane in RN spacetime with that in Schwarzschild and Newtonian case using $V_{\rm RN}$. Solid and long-dashed curves in all the figures denote Newtonian and Schwarzschild cases, respectively. Short-dashed and dotted curves in figures 14a,b,c correspond to $r_Q/r_s = (0.5, 0.1)$, $r_Q/r_s = (1.116, 1.8)$ and $r_Q/r_s = (2.1, 2.5)$, respectively. Other parameters are identical to that in Fig. 13.  
 }
\label{Fig14}
\end{figure}

In figures 13 and 14, we show the trajectory profiles of a test particle for parabolic like orbit using $V_{\rm JNW}$ and $V_{\rm RN}$ corresponding to JNW and RN geometries for various values of $\gamma$ and $r_Q$, respectively, with a comparison to those in Newtonian and Schwarzschild cases. For both the JNW and RN geometries, the test particle starts from an arbitrary source $S_{(x,y)}$ with an arbitrary fixed initial velocity. We choose the initial velocity $v_{\rm in} \equiv v_y = - 0.1732\, c$ in our studies, for all corresponding values of $\gamma$ and $r_Q$ and also for the Newtonian case. It is found that with the preferred orbital parameters chosen here, no proper well defined orbits would be produced for values of $\gamma < 0.2$ for JNW geometry as the particle will simply plunge into the naked singularities (Fig. 13a). In Table 1, we furnish the corresponding values of the particle's {\it least distance of approach} ${\beta}_p$ along with their transit time $T_{\rm tr}$, i.e., the time taken by the particle to traverse the distance from $S_{(x,y)}$ to the locations of their corresponding ${\beta}_p$, around both JNW and RN geometries. ${\beta}_p \vert_{\rm NT}$ and 
$T_{\rm tr} \vert_{\rm NT}$ correspond to the Newtonian case. It is found that, corresponding to JNW geometry, the 
magnitude of ${\beta}_p$ continuously decreases from that of the corresponding value in Schwarzschild geometry as one departs more from the Schwarzschild BH solution, i.e., with the decrease of $\gamma$. This is in contrary to the situation in RN geometry, where ${\beta}_p$ always increases as one departs from the Schwarzschild BH solution, i.e., with the increase of $r_Q$ having naked singularities. Equivalently, this implies that in JNW geometry if the test particle starts from a fixed source with a fixed initial velocity, then, as one departs more from the Schwarzschild BH solution, the bending angle corresponding to a parabolic like trajectory steadily decreases. On the contrary, for RN geometry as one departs more from the Schwarzschild BH solution, exactly opposite thing occurs. This opposite behavior of the particle trajectory profiles corresponding to JNW and RN geometries can be attributed to the opposite nature of the variation of the terms $\left(1-\frac{2GM}{\gamma c^2 r}\right)^\gamma$ and $\left(1- \frac{2r_s}{r} + \frac{r^2_Q}{r^2} \right)$ in the corresponding metrics of JNW and RN geometries, as one departs from the Schwarzschild solution. From these two terms, one can see that to the first order they resemble Schwarzschild geometry. However, if we analyze these terms to the second order, it reveals that the net effect of the decrease in the value of $\gamma$ as one departs from the Schwarzschild solution corresponding to JNW geometry, is to effectively diminish the curvature effect of gravity. On the contrary, corresponding to RN geometry, the net effect of the increase in the value of $r_Q$ as one departs from the Schwarzschild solution is to effectively enhance the curvature effect of gravity. Owing to which, for JNW geometry, the 
bending angle for a parabolic like trajectory steadily decreases as one departs from the Schwarzschild BH solution, while the bending angle for a parabolic like trajectory steadily increases in RN geometry as one departs more from the Schwarzschild solution. Similar behavior can also be seen for photon trajectories in the presence JNW and RN geometries. Nonetheless, for both these geometries, the corresponding transit time $T_{\rm tr}$ always increases as one departs from the Schwarzschild BH solution. 

The change in the initial value of the orbital parameters like $v_{\rm in}$ or the location of the source $S_{(x,y)}$ do not fundamentally alter the nature of parabolic like particle trajectories corresponding to both JNW and RN geometries; the qualitative nature of the variation of ${\beta}_p$ or the bending angle and $T_{\rm tr}$ with $\gamma$ and $r_Q$ remains independent of the choice in the value of $v_{\rm in}$ or $S_{(x,y)}$, and is similar to that depicted in Table 1. Nonetheless, with the decrease in the magnitude of $\vert v_{\rm in} \vert$, corresponding to all values of $\gamma$ and $r_Q$, there is a steady increase in the corresponding values of ${\beta}_p$ or the bending angle. And with a further decrease in the value $\vert v_{\rm in} \vert$, the unbound parabolic like particle orbits tend to become eventually bound or elliptical in nature. Moreover, with the decrease in the value of $v_{\rm in}$ in JNW geometry, well defined orbits are formed only for $\gamma > 0.2$. On the other hand, with the decrease in the distance of the location of the source (in the y-direction) from the central gravitating mass, here too, corresponding to all $\gamma$ and $r_Q$, there is a marginal increase of ${\beta}_p$ or the bending angle. 
\begin{table}
\large
\centerline{\large Table 1}
\centerline{\large $S_{(x,y)} = (60, 400) \, r_s,  \, v_x = 0, \, v_y = - 0.1732 \, c$ }
\centerline{\large ${\beta}_p \vert_{\rm NT} = 36.922 \, r_s$, \, \, \,   $T_{\rm tr} \vert_{\rm NT} = 2149.8 \, {r_s}/c$ }
\begin{center}
\begin{tabular}{ccccccccccccc}
\hline
\hline

\noalign{\vskip 2mm} 
$\rm JNW$ &  ${\beta}_p \, (r_s)$ &  $T_{\rm tr} \left({r_s}/c \right)$ & $\rm RN$ &  ${\beta}_p \, (r_s)$  &  $T_{\rm tr} \left({r_s}/c \right)$ \\
\hline
\noalign{\vskip 2mm} 
$\gamma = 1.0$  &   39.027 & 2163 & $r_Q = 0.0 \, r_s$  &  39.027 & 2163 \\
\hline
\noalign{\vskip 2mm}
$\gamma = 0.95$  &  38.891 & 2163.4 & $r_Q = 0.5 \, r_s$  &  39.081 & 2163.1 \\
\hline
\noalign{\vskip 2mm}
$\gamma = 0.8$  &  38.375 &  2165.1 & $r_Q = 1.0 \, r_s$  &  39.242 & 2163.3 \\
\hline
\noalign{\vskip 2mm}
$\gamma = 0.7$  &  37.895 & 2166.6 & $r_Q = 1.116 \, r_s$  &  39.295 & 2163.4 \\
\hline
\noalign{\vskip 2mm}
$\gamma =  0.5$  &  36.273  &  2171.9 & $r_Q = 1.8 \, r_s$  &  39.724 & 2164\\
\hline
\noalign{\vskip 2mm}
$\gamma  = 0.4$  &  34.721  &  2177.2  & $r_Q = 2.1 \, r_s$  &  39.976 & 2164.4 \\
\hline
\noalign{\vskip 2mm}
$\gamma  = 0.3$  &  31.745 &   2188.5 & $r_Q = 2.5 \, r_s$  &  40.372 & 2164.9 \\
\hline
\noalign{\vskip 2mm}
$\gamma  = 0.2$  &  20.79 &   2272.8  &  & &  &  & &  &  & & \\
\hline
\hline
\end{tabular}
\end{center}
\end{table}

\begin{table}
\large
\centerline{\large Table 2}
\centerline{\large $S_{(x,y)} = (60, 400) \, r_s, \, O_{(x,y)} = (-400.04, -282.25) \, r_s, \, v_y = -0.1732 \, c$ }
\centerline{\large ${\beta}_p \vert_{\rm NT} = 36.922 \, r_s$,  \, \, \,   $T_{\rm tr} \vert_{\rm NT} = 2149.8 \, {r_s}/c$,  \, \, \, $T_{\rm tot} \vert_{\rm NT} = 4799.5 \, {r_s}/c$ }
\begin{center}
\begin{tabular}{cccccccccccccccccccccccc}
\hline
\hline

\noalign{\vskip 2mm} 
$\rm JNW$ &  $v_x \,(c)$ & ${\beta}_p \, (r_s)$ &  $T_{\rm tr} \left({r_s}/c \right)$ &  $T_{\rm tot} \left({r_s}/c \right)$ & $\rm RN$ &  $v_x \, (c)$ & ${\beta}_p \, (r_s)$ &  $T_{\rm tr} \left({r_s}/c \right)$ & $T_{\rm tot} \left({r_s}/c \right)$ \\
\hline
\noalign{\vskip 2mm} 
$\gamma = 1.0$ & −6.11E-4 &  37.737 & 2160.4  & 4820  & $r_Q = 0.0 \, r_s$  & −6.11E-4 & 37.737 & 2160.4 & 4820 \\
\hline
\noalign{\vskip 2mm}
$\gamma = 0.95$  & -3.64E-4 & 38.123 & 2161.9 & 4823 & $r_Q = 0.5 \, r_s$ & -7.4E-4 &  37.522 & 2159.9 & 4819 \\
\hline
\noalign{\vskip 2mm}
$\gamma = 0.8$ & 5.41E-4  &  39.527 &  2167.3 & 4834.5 & $r_Q = 1.0 \, r_s$ & -1.137E-3 &  36.857  &  2158.4 & 4816 \\
\hline
\noalign{\vskip 2mm}
$\gamma = 0.7$ & 1.327E-3 &  40.747  &  2172.1 & 4844 & $r_Q = 1.116 \, r_s$ & -1.27E-3 &  36.633 &  2157.9 & 4815 \\
\hline
\noalign{\vskip 2mm}
$\gamma =  0.5$ & 3.687E-3   &  44.391 &  2186.3 & 4873.5 & $r_Q = 1.8 \, r_s$ & -2.435E-3 & 34.677 & 2153.4  & 4805.5 \\
\hline
\noalign{\vskip 2mm}
$\gamma  = 0.4$ & 5.598E-3 &  47.324  &  2197.80 & 4897 & $r_Q = 2.1 \, r_s$ & -3.202E-3 &  33.387  & 2150.2 & 4799 \\
\hline
\noalign{\vskip 2mm}
$\gamma  = 0.3$ & 8.552E-3 &  51.828 &  2215.4 & 4933.5 & $r_Q = 2.5 \, r_s$ & -4.596E-3 &  31.044  & 2144.3 & 4786.5 \\
\hline
\noalign{\vskip 2mm}
$\gamma  = 0.2$ & 1.387E-2 &  59.839  &  2247  & 4998.5  & &  &  & &  &  & & \\
\hline
\noalign{\vskip 2mm}
$\gamma  = 0.1$ & 2.2735E-2 &  79.534  &  2324.6  & 5158  & &  &  & &  &  & & \\
\hline
\hline
\end{tabular}
\end{center}
\end{table}

In the next scenario, we study the trajectory profiles of a test particle for parabolic like orbit keeping both the locations of the source $S_{(x,y)}$ and the observer $O_{(x,y)}$ fixed, unlike the previous case where only the location of the source was fixed. Owing to which the test particle needs to start from an arbitrary source $S_{(x,y)}$ obliquely with different initial velocities, corresponding to different values of $\gamma$ and $r_Q$, in order to reach the fixed location of the observer $O_{(x,y)}$. Here we choose the similar values of $S_{(x,y)}$ and the initial velocities for $v_y$ like that in the previous case for all values of $\gamma$ and $r_Q$, however, with different initial velocities for $v_x$. In the $x-y$ coordinate plane, the central object is considered to be situated at $(0,0)$. In Table 2 we display the computed values of ${\beta}_p$, $T_{\rm tr}$, and the time taken by the particle to traverse the distance from $S_{(x,y)}$ to $O_{(x,y)}$ $\left(T_{\rm tot}\right)$, for all values $\gamma$ and $r_Q$ corresponding to JNW and RN geometries. The corresponding different values for $v_x$ for JNW and RN geometries are shown in Table 2. $T_{\rm tot} \vert_{\rm NT}$ in Table 2 denotes the total time traverse in the Newtonian case. Interestingly it is found from Table 2 that in the JNW geometry the magnitude of ${\beta}_p$ or equivalently the bending angle continuously increases as one departs more from the Schwarzschild BH solution, while in RN geometry ${\beta}_p$ or equivalently the bending angle continuously decreases as one departs from the Schwarzschild BH solution, even attaining less value then that in the Newtonian case for naked singularities. This is in sharp contrast to the scenario in the previous case, where exactly opposite behavior 
prevails. On the other hand, both the magnitudes of $T_{\rm tr}$ and $T_{\rm tot}$ steadily increase with the decrease  of $\gamma$ in JNW geometry. On the contrary, in RN geometry, both $T_{\rm tr}$ and $T_{\rm tot}$ continuously decrease as one departs more from the Schwarzschild BH solution; the time taken by the particle to reach $O_{(x,y)}$ for naked singularities even becoming less as compared to the Newtonian case .  

The kind of trajectories those we have studied here are important in probing the gravitational field around the naked singularities in strong field regime. Such studies can be exploited to evaluate gravitational bending of light and perihelion precession of test particles which may offer to distinguish between naked singularity solutions and BHs observationally.

\section{Accretion disk}

In this section we analyze a simplistic accretion flow system in JNW and RN geometries using their respective analogous potentials described in Eqs. (12) and (35). The JNW and RN analogous potentials quite precisely mimic the corresponding GR features in their entirety. For this we consider the simple model of a stationary, geometrically thin and optically thick Keplerian accretion disk, also called the standard accretion disk model of Shakura and Sunyaev [27]. Although this analytic model has been initially developed in context to Newtonian gravitational potential, however, later modifications using PNPs corresponding to other relativistic geometries have also been accomplished (e.g., [9]), in order to model geometrically thin and optically thick accretion flow studies around BHs/compact objects. The two most important results obtained from the standard accretion disk model are the amount of radiative flux ($F_{\rm rad}$) and the 
luminosity ($L_{\rm rad}$) generated from the optically thick Keplerian accretion disk, whose expressions are given by (see [9])
\begin{eqnarray}
F_{\rm rad} = \frac{{\mathcal{Q}}^{+}}{2} = \left( - {\dot M} \right) \, 
\left(- \frac{d\Omega^K}{dr} \right) \, \left(\lambda^K - \lambda^K_{\rm in} \right) 
\label{40}
\end{eqnarray}  
and
\begin{eqnarray}
L_{\rm rad} = 2 \int^{\infty}_{r_{\rm in}}  \left(- {F}_{\rm rad} \right)  \, 2 \pi r \, dr \, , 
\label{41}
\end{eqnarray}
respectively. ${\mathcal{Q}}^{+}$ is the total heat generated due to turbulent viscosity in the column of the disk, ${\dot M}$ is the usual mass accretion rate. $\Omega^K$ and $\lambda^K$ are Keplerian angular velocity and Keplerian angular momentum, respectively. $r_{\rm in}$ is the radius of the inner edge of the disk, which in this case would be the radius of the marginally stable orbit $r_{\rm ms}$. For a Keplerian accretion flow, we use the conditions $\dot r =0$ and 
$\dot{\Omega} = \Omega^K$ in the expressions for potential $V$ in Eqs. (12) and (37), corresponding to JNW and RN geometries, respectively. Using the relations $\Omega^K = \left(\frac{1}{r} \frac{dV}{dr}\right)^{1/2}$ and 
$\lambda^K = \left(r^3 \frac{dV}{dr}\right)^{1/2}$ corresponding to Keplerian accretion flow, we eventually obtain the relations for $\Omega^K$ and $\lambda^K$ corresponding to Keplerian accretion flow in JNW and RN spacetimes, which is given by 
\begin{eqnarray}
\Omega^K \vert_{\rm JNW} = \frac{\left(1-\frac{2r_s}{\gamma r}\right)^{2\gamma -1}}{r^2} \, \sqrt{\frac{c^2 r r_s (\gamma r)^\gamma}{(\gamma r - 2r_s)^\gamma - (2\gamma -1) (\gamma r - 2r_s)^{\gamma -1} r_s }  } \, ,
\label{42}
\end{eqnarray}

\begin{eqnarray}
\lambda^K \vert_{\rm JNW} = \sqrt{\frac{c^2 r r_s (\gamma r)^\gamma}{(\gamma r - 2r_s)^\gamma - (2\gamma -1) (\gamma r - 2r_s)^{\gamma -1} r_s }  } \, ,
\label{43}
\end{eqnarray}

\begin{eqnarray}
\Omega^K \vert_{\rm RN} =  \frac{r - 2r_s +\frac{r^2_Q}{r} }{r^2} \sqrt{\frac{GM - \frac{c^{2} r_{Q}^{2}}{r}}{r-3r_{s} + \frac{2r_{Q}^{2}}{r}} } \,  
\label{44}
\end{eqnarray}
and
\begin{eqnarray}
\lambda^K \vert_{\rm RN} =  r \sqrt{\frac{GM - \frac{c^{2} r_{Q}^{2}}{r}}{r-3r_{s} +  \frac{2r_{Q}^2}{r}} } \, ,
\label{45}
\end{eqnarray}

\begin{figure}
\begin{center}
\includegraphics[width = 1.0\textwidth,angle=0]{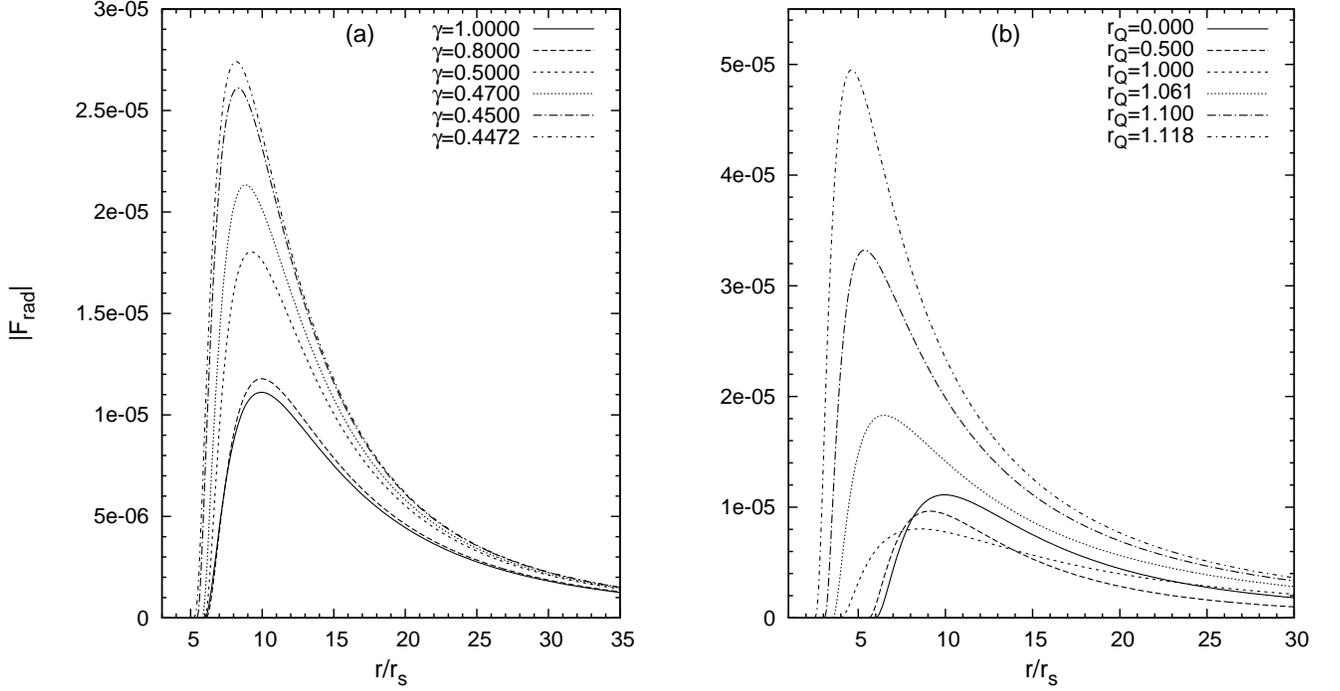}
\end{center}
\caption{Variation of the radiative flux $\vert F_{\rm rad} \vert$ generated from a geometrically thin and optically thick Keplerian accretion disk with radial distance $r$ for various $\gamma$ and $r_Q$ corresponding to Keplerian accretion flow around JNW and RN geometry, respectively. Figure 15a is for various $\gamma$ over 
$1 \leq \gamma \leq 0.4472$ corresponding to $r_{\rm ms}$ of outer loci. Similarly Fig. 15b is for various $r_Q$ in the range
$0 \leq r_Q \leq 1.118 \, r_s$ corresponding to $r_{\rm ms}$ of outer loci.  We considered $\dot M =1$, $G=M=C=1$. 
 }
\label{Fig15}
\end{figure}

It should be noted that corresponding to JNW and RN geometry, for $0.4472 \, \lsim \, \gamma < 0.5$ and $\sqrt{9/8} \, r_s \, < r_Q \, \lsim \, 1.118 \, r_s$ respectively, we will have two values of $r_{\rm in}$ corresponding to two values of $r_{\rm ms}$ represented by outer $\left(r_{\rm ms} \vert_{\rm out} \right)$ and inner 
$\left(r_{\rm ms} \vert_{\rm in} \right)$ loci, respectively. The orbit of $r_{\rm ms} \vert_{\rm out}$ is separated from 
$r_{\rm ms} \vert_{\rm in}$ by a {\it zone of instability}. The formulation of a geometrically thin Keplerian accretion disk model in the framework of Shakura and Sunyaev model would only be compatible with $r_{\rm ms} \vert_{\rm out}$ which will then be the inner edge ($r_{\rm in}$) of the accretion disk. Using Eqs. (42-45), we compute the magnitude of the 
radiative flux $\vert F_{\rm rad} \vert$ for different values of $\gamma$ and $r_Q$ corresponding to JNW and RN geometries, respectively, using appropriate values of $r_{\rm in} \left(\equiv r_{\rm ms} \vert_{\rm out} \right)$ and show the radial profiles of $\vert F_{\rm rad} \vert$ in Fig. 15. In RN geometry, with the increase of $r_Q$ up to $r_Q = r_s$, there is a steady decrease in the magnitude of the peak of the $\vert F_{\rm rad} \vert$ profiles, which is in sharp contrast to the profiles for naked singularities corresponding to both RN and JNW geometries with $r_{\rm ms}$ as outer loci.

\begin{table}
\large
\centerline{\large Table 3}
\centerline{Radiative efficiency}
\begin{center}
\begin{tabular}{ccccccccccccc}
\hline
\hline
\noalign{\vskip 2mm} 
$\rm JNW$ &  $\eta$ & $\rm RN$ &  $\eta$ \\
\hline
\noalign{\vskip 2mm} 
$\gamma = 1.0$  &  $\sim 0.056$ &   $r_Q = 0.0 \, r_s$  &  $\sim 0.056$ \\
\hline
\noalign{\vskip 2mm}
$\gamma = 0.8$  &  $\sim 0.057$ &  $r_Q = 0.5 \, r_s$  &  $\sim 0.019$  \\
\hline
\noalign{\vskip 2mm}
$\gamma = 0.5$  &  $\sim 0.067$ &   $r_Q = 0.8 \, r_s$ & $\sim 0.046$  \\
\hline
\noalign{\vskip 2mm}
 $\gamma = 0.48$ &   $\sim 0.069$ & $r_Q = 1.0 \, r_s$  & $\sim 0.12$  \\
\hline
\noalign{\vskip 2mm}
$\gamma = 0.46$  & $\sim 0.073$ & $r_Q = 1.08 \, r_s$ & $\sim 0.163$  \\
\hline
\noalign{\vskip 2mm}
$\gamma =  0.45$  &  $\sim 0.076$  &  $r_Q = 1.1 \, r_s$  &  $\sim 0.18$  \\
\hline
\noalign{\vskip 2mm}
$\gamma  = 0.4472$  &  $\sim 0.078$  &  $r_Q = 1.118 \, r_s$  &  $\sim 0.195$  \\
\hline
\hline
\end{tabular}
\end{center}
\end{table}

In Table 3, we furnish the radiative efficiency $\eta \left(= {L_{\rm rad}}/{\dot M c^2}\right)$
for relevant values of $\gamma$ and $r_Q$ with $r_{\rm ms}$ corresponding to $\left(r_{\rm ms} \vert_{\rm out} \right)$. For both JNW and RN naked singularity geometries, as one deviates more from the Schwarzschild solution the radiative efficiency $\eta$ continuously increases. This nature of the variation of the radiative efficiency $\eta$ for different $\gamma$ and $r_Q$ can be well accounted from the radiative flux profiles in Fig. 15. 

Note that while analyzing the Keplerian accretion flow problem in JNW and RN geometries as shown in Fig. 15 and Table 3, we have restricted up to $\gamma = 0.4472$ and $r_Q = 1.118 \, r_s$ as for $\gamma < 0.4472$ or $r_Q > 1.118 \, r_s$, no stable Keplerian accretion disk will be formed. 

\section{Discussion}

Naked singularities may occur as an alternative end state of gravitational collapse instead of BH solutions, where a significant departure from BH solutions could occur through a permeating scalar field or spontaneous scalarization due to continuous matter distribution [28] or even through very strong electromagnetic field. If naked singularities are indeed present in the nature, it is important to figure out the observationally distinguishing features, at least in principle at this stage, of naked singularities in comparison to black holes. The general relativity so far has been tested in solar system measurements as well as by accurate radio observations of binary pulsars. However, in all such cases, the gravitational field is weak and it appears from the detailed studies over the last few decades that naked singularities cannot be discriminated from black holes from weak field observations; one has to look for strong field tests instead. 

The characteristics of the radiation emitted from the accretion disk are supposed to provide useful details about the spacetime geometry around the compact object. We studied a very simple accretion model exploiting Newtonian like analogous potential of the corresponding naked singularity geometries. The very premise with which the Newtonian like analogous potential of the corresponding static GR geometries have been derived in the present work ensures reproduction of identical or near identical geodesic equations of motion. Not only the orbits like marginally stable or marginally bound are exactly reproduced, dynamical profiles like conserved angular momentum, conserved energy, the temporal features like angular and epicyclic frequencies are reproduced with precise accuracy. Most importantly, the adopted method guarantees the replication of the orbital trajectory of test particle motion accurately, consequently, reproducing the experimentally tested GR effects like perihelion advancement, gravitational bending of light or gravitational time delay with precise accuracy. The generality of the procedure then ensures that not only static GR geometries with event horizons can be comprehensively mimicked through this kind of potential; GR features corresponding to naked singularities can also be reproduced comprehensively with precise accuracy and can be used to analyze relevant astrophysical processes in strong field gravity around corresponding naked singularities. 

It is found from the present analysis that accretion disk properties around naked singularities show clear notable differences, with geometrically thin Keplerian disk more luminous than that around equivalent BHs. Out of the two possible locations of $r_{\rm ms} (r_{\rm ms} \vert_{\rm in}$ and $r_{\rm ms} \vert_{\rm out})$ for certain range in the values of $\gamma$ and $r_Q$ corresponding to naked singularity solutions, the inner edge of the Keplerian accretion disk around a naked singularity would be located at $r_{\rm ms} \vert_{\rm out}$. However,  gaseous particles reaching $r_{\rm ms} \vert_{\rm out}$ would then simply plunge either to reach the singularity or $r_{\rm ms} \vert_{\rm in}$ where stable circular orbits occur. Nonetheless, for non-Keplerian accretion flow around a naked singularity, this inner region ($r<r_{\rm ms} \vert_{\rm out}$) may become very important where the dynamical behavior of the accretion disk may be significantly different than that around equivalent BH solutions, consequently the corresponding spectrum from the accretion flow around a naked singularity may show certain distinctive features as compared to the similar accretion flow around BH solutions. Thus X-ray binaries or active galactic nuclei would be the most likely regions where distinguishable differences in the observational features between BHs and naked singularities, can, in principle, be made.

An important point is that, whether the derived potentials contain the naked singularity features of the concerned spacetime geometries. Let us first consider the case of JNW solution. The Ricci, the Kretschmann, and the Weyl scalars for the JNW are known to diverge at singularities. The derived modified Newtonian analogous potential corresponding to JNW metric is also found to diverge at $r = 2r_s/\gamma$ but the radial velocity remains finite as follows from the Eq. (16) and radial infall to the singularity is admissible. Being a globally naked singularity solution, there exists a future directed causal curve with one end on the singularity and the other end on future null infinity for the JNW geometry in the GR treatment. After recovering test particle mass (m), for radial motion the Eq. (16) reads $\frac{dr}{dt} = f(r)^\gamma \, \sqrt{2 \, E_{\rm GN}^{'} - m^2 c^2 \, \left(f(r)^\gamma -1 \right) }$ which for photons ($m=0$) becomes $\frac{dr}{dt} = f(r)^\gamma \, \sqrt{2 \, E_{\rm GN}^{''}}$  \, [where $E_{GN}^{'}$, the constant of motion for test particle of mass $m$ is the conserved energy (not the specific energy), and $E_{GN}^{''}$ is the equivalent term corresponding to photon]. Therefore we get the same expression of radial velocity as given by the GR treatment in the low energy limit. For an observer at a finite distance R, the solution of outgoing null geodesic from the singularity is 
\begin{eqnarray}
\underset{\epsilon \rightarrow 0}{lim} \int^{R}_{2r_{s}/\gamma+\epsilon} f^{-\gamma}dr < R \; \underset{\epsilon \rightarrow 0}{lim} \int^{R}_{2r_{s}/\gamma+\epsilon} \frac{dr}{\left(r-2r_{s}/\gamma \right)^{\gamma}} 
= R \frac{\left(R-2r_{s}/\gamma \right)^{1-\gamma}}{1-\gamma}
\end{eqnarray}
which is finite for $\gamma < 1$, where $\epsilon$ is a small perturbation along radial distance $r$. Hence the modified Newtonian analogous potential corresponding to JNW metric admits outgoing null geodesic from the singularity (i.e. the singularity is visible to external observers) and thus the derived potential contains the globally naked singularity features of the original spacetime. The same argument can be extended for the RN case with $r_Q > r_s$.

\section{Conclusion}

In this work, we have formulated a generic Newtonian like analogous potential corresponding to static spherically symmetric general GR spacetime, and subsequently derived proper Newtonian like analogous potential for JNW and RN spacetimes. The derived PNPs found to reproduce the entire GR features with precise accuracy. We also studied orbital dynamics around these two geometries extensively in the modified Newtonian analogue, including the detailed analysis of their corresponding test particle trajectories.

Our findings employing the modified Newtonian analogous potentials for the naked singularity solutions show that as one departs more from Schwarzschild solution i.e. with the increase in the value of $\gamma$ and $r_Q$, the nature of the test particle dynamics along circular orbit in JNW and RN spacetimes tends to show altogether different behavior with distinctive traits as compared to the nature of particle dynamics in the Schwarzschild geometry. Interestingly we found  two values of $r_{\rm ms}$ for a certain range in the value of $\gamma$  ($0.4472 \, \lsim \, \gamma < 0.5$) and $r_Q$ ($\sqrt{9/8} \, r_s < r_Q \, \lsim \, 1.118 \, r_s$), which are unique features of JNW and RN geometries. For values of $\Gamma < 0.4472$ and $r_Q > 1.118 \, r_s$, the particle in circular trajectory would not have last stable 
orbit, consequently if the particle motion is perturbed in the inner region corresponding to JNW and RN geometries with these values of $\gamma$ and $r_Q$, the particle would eventually plunge into the naked singularities. 

The magnitude of the apsidal precession in the JNW geometry increases continuously as one departs more from the Schwarzschild BH solution, irrespective of the orbital parameters up to the value of $\gamma \sim 0.45$, beyond which, the particle trajectory does not produce well defined orbits. On the contrary, for the RN geometry, the magnitude of the apsidal precession decreases as one departs from the Schwarzschild BH solution, however beyond a certain value of $r_Q$ depending on the orbital parameters, the magnitude of the apsidal precession increases, with the particle orbit showing retrograde precession. For particular value of $r_Q$ ($r_Q \sim 1.68 \, r_s$) in RN geometry, the value of apsidal precession becomes almost independent of orbital parameters for the elliptical orbits. If the test particle starts from a fixed source with a fixed initial velocity irrespective of the location of the source and magnitude of the initial velocity, as one departs more from the Schwarzschild BH solution, the bending angle of particle's parabolic like orbital trajectory in JNW geometry steadily decreases, even attaining less value then that of the Newtonian case, while there is a steady increase in the the bending angle in RN geometry. On the contrary, if the test particle starts with different initial velocity corresponding to different values of $\gamma$ or $r_Q$ from a fixed source to reach the fixed location of the observer, as one departs more from the Schwarzschild BH solution, the bending angle of particle's parabolic like orbital trajectory along with the total time taken by the particle to reach the observer corresponding to JNW geometry increases, in contrast to the case in RN geometry where the bending angle as well as the time taken by the particle to reach the observer steadily decreases, even becoming less as compared to that of the Newtonian case. One can then infer that the gravitational bending of light would also show similar kind of behavior in the presence of these geometries having naked singularities. 

We applied the Newtonian like analogous potentials to model a simple geometrically thin and optically thick Keplerian accretion disk in the presence of JNW and RN geometries. We found that the radiative efficiencies of Keplerian accretion disk around both JNW and RN naked singularities are always higher than that around Schwarzschild geometry. Although our analysis has been performed for a simplistic geometrically thin Keplerian accretion disk, however, the nature of the variation of the radiative efficiency with $\gamma$ or $r_Q$ for JNW and RN geometries, likely to be remain similar even for complex accretion flow processes in the presence of these geometries. More extensive study of accretion flow processes including the detailed modeling of geometrically thick and/or advective accretion flow in these geometries would be pursued in a later work. The accreting system in the strong field regime, thus, appears to provide a natural laboratory to ascertain the presence of either naked singularities or BHs.  

Here it is worthwhile to mention that in reality BHs probably have almost no electric charge because a charged BH is expected to quickly neutralize by attracting charge of the opposite sign [29]. Hence RN BHs/naked singularities might not have much astrophysical relevance. On the other hand Price theorem [30] suggests that the asymptotically flat scalar fields around a BH should radiate away quickly leaving only its constant asymptotic value and the Schwarzschild BH, thereby raising doubt on the physical reality of the JNW spacetime. However, the Price theorem is not strictly applicable to JNW metric as the metric has no horizon and the scalar field diverges at curvature singularity [31]. Here it is worthwhile to mention that there are several claims in the literature for formation of naked singularity in generic gravitationally spherical collapse of an inhomogeneous dust ball but whether the JNW solution will occur in a generic gravitational collapse is not known yet. It would be interesting if the present approach can be extended to the recently found stationary BH solutions with long-lived (complex) scalar field [32] to study the influence of scalar field in accretion related phenomena. 

\section*{Acknowledgments}
The authors would like to thank the anonymous reviewers and the adjudicator for insightful comments and very useful suggestions that helped us to improve the manuscript.


\end{document}